\documentclass[11pt]{article}
\usepackage[latin1]{inputenc}
\usepackage{amssymb}
\usepackage{hyperref}
\usepackage{amsmath}
\usepackage{latexsym}
\usepackage{times}
\usepackage{cite}
\usepackage{amssymb}
\usepackage{amsfonts}
\usepackage{amscd}
\usepackage{xcolor}
\usepackage{amstext,amsmath,amssymb,amsfonts}
\newtheorem{theorem}{Theorem}

\newtheorem{corollary}[theorem]{Corollary}

\newtheorem{definition}[theorem]{Definition}

\newtheorem{lemma}[theorem]{Lemma}

\newtheorem{proposition}[theorem]{Proposition}
\newtheorem{remark}[theorem]{Remark}

\newcommand{\beq}{\begin{eqnarray}}
\newcommand{\eeq}{\end{eqnarray}}
\newcommand{\beqs}{\begin{eqnarray*}}
	\newcommand{\eeqs}{\end{eqnarray*}}
\newcommand{\bpro}{\begin{pro}}
	\newcommand{\epro}{\end{pro}}
\newcommand{\blem}{\begin{lem}}
	\newcommand{\elem}{\end{lem}}
\newcommand{\bdfn}{\begin{dfn}}
	\newcommand{\edfn}{\end{dfn}}
\newcommand{\bcor}{\begin{cor}}
	\newcommand{\ecor}{\end{cor}}
\newcommand{\bthm}{\begin{thm}}
	\newcommand{\ethm}{\end{thm}}
\newcommand{\bex}{\begin{ex}}
	\newcommand{\eex}{\end{ex}}
\newcommand{\brmk}{\begin{rmk}}
	\newcommand{\ermk}{\end{rmk}}
\newcommand{\bpr}{\begin{pr}}
	\newcommand{\epr}{\end{pr}}
\newcommand{\benum}{\begin{enumerate}}
	\newcommand{\eenum}{\end{enumerate}}
\newcommand{\bitem}{\begin{itemize}}
	\newcommand{\eitem}{\end{itemize}}

\newcommand{\cqfd}{\hfill{\square}}
\chardef\bslash=`\\
\numberwithin{equation}{section}
\numberwithin{table}{section}
\numberwithin{theorem}{section}

\begin{document}
	\begin{flushright}
		ICMPA-MPA/2018
	\end{flushright}
	\begin{center}
		{\Large {Geometry and probability on the noncommutative\\ 2-torus in a magnetic field} }\\
		\vspace{0,5cm}
		Mahouton Norbert Hounkonnou$^{*}$ and Fridolin Melong\\
		\vspace{0.5cm}
		{\em International Chair in Mathematical Physics
			and Applications}
		{\em (ICMPA-UNESCO Chair) }\\
		{\em University of Abomey-Calavi,}
		{\em 072 B.P. 50 Cotonou, Republic of Benin}\\	
{\em E-mails: norbert.hounkonnou@cipma.uac.bj, (with copy to hounkonnou@yahoo.fr);}
{\em fridomelong@gmail.com}
	\end{center}
	\today
	
	\vspace{0.5 cm}
	{\it Abstract}
	
In this work, we describe the geometric and probabilistic   properties of a noncommutative $2-$ torus in a magnetic field. We study the volume  invariance, integrated scalar curvature and volume form by using the method of perturbation by inner derivation of the magnetic Laplacian  in the noncommutative $2-$torus. Then,  we analyze the magnetic stochastic process  describing the motion of a particle subject to  a uniform magnetic field on the noncommutative $2-$ torus,  derive and discuss the related main properties.
\vspace{0.5 cm}\\
\vspace{0.5 cm}
\tableofcontents
\section{Introduction}
Noncommutative geometry (NG), initiated by Alain Connes  \cite{Con},  is an exciting dynamic  research area of mathematics with applications in physics, greatly developed over the last decades (see  \cite{GHM},\cite{HMG} and references therein). The Gelfand Naimark theorem \cite{KW} which gives an anti-equivalence between the category of locally compact Hausdorff spaces and the one of commutative $C^*$-algebra $\mathcal{A}$, is the starting point of noncommutative topology \cite{K} ,\cite{KW}. The correspondence is given by the map $ X\longrightarrow C_0(X),$ where $C_0$ is the algebra of continuous complex valued functions \cite{S}. Thus,  one thinks of noncommutative $C^*$-algebras as noncommutative topological spaces and tries to apply topological methods to understand them \cite{S}.

Sakamoto and Tanimura in \cite{ST}, extended the Fourier analysis for the noncommutative $n-$ torus in a magnetic field. In particular, they studied the solutions of the Schr{\"o}dinger equation in a uniform magnetic field for  the noncommutative $n-$ torus. They  defined a magnetic algebra and showed that,  when the space of functions is an irreductible representation space of this  algebra, it characterizes the quantum mechanics in the magnetic torus. In addition,  they solved the eigenvalue problem of the Laplacian for the magnetic torus, and   provided simple forms for all eigenfunctions.

In \cite{CGS},  Chakraborty,  Goswami, and  Sinha  studied, in a noncommutative $2-$torus, the effect of perturbation by inner derivation
on the  quantum stochastic process and geometric parameters
like the volume and the scalar curvature. Their cohomological calculations
showed that the above perturbation produces new spectral triples.  They also obtained,
for the Weyl $C^{*}$-algebra $\mathcal{A}_{\theta},$(algebra of noncommutative $2$-torus developed in \cite{R}), the Laplacian associated with a natural stochastic
process  and computed the associated volume form.

The purpose of our  paper is to elucidate the geometric and stochastic properties of a  noncommutative  $2$-torus in a uniform magnetic field. Especially, we define a $2$-torus noncommutative magnetic Laplacian, and derive the related magnetic quantum stochastic differential equation on the line of the construction  given in \cite{CGS}, \cite{GS}, \cite{GSK}, and \cite{Sav}. Further, from the Weyl asymptotics for the constructed  $C^{*}-$ algebra $\mathcal{D}_{\theta}$, we obtain the volume and integrated scalar curvature for the $2$-torus, and investigate their invariance under perturbation by inner derivation of the noncommutative magnetic Laplacian. The  volume form invariance is also analyzed. Finally, using the magnetic position and  momentum operators in a noncommutative $2d$ space, we establish the associated magnetic  quantum stochastic differential equation, and discuss the  properties of its solution.

The paper is  organized as follows. In section $2$, we briefly recall the basic definitions and properties of the magnetic Laplacian. In section $3,$  we construct and discuss the associated magnetic quantum stochastic diffusion equation. In section $4$, we introduce the volume, the integrated scalar curvature using Weyl asymptotics for the  $C^{*}$-algebra of the noncommutative $2$- torus, and study their invariance under the perturbation of the noncommutative magnetic Laplacian. {In section $5,$ we investigate the invariance of the volume form under the above perturbation  and explore the spectral triple on $\mathcal{D}_{\theta}$}.  The magnetic quantum stochastic differential equation in the case of  a noncommutative $2d$-dimensional space is investigated and analyzed in Section ${6}.$ Section ${7}$ is devoted to some  concluding remarks.
\section{Basic  properties}
We consider   a finite dimensional real vector space $V$ equipped with a positive defined inner product $\langle \, , \, \rangle$.
Let $\theta$ be  an irrational number in $[0,1]$.  The irrational rotation $C^{*}-$ algebra $\mathcal{D}_{\theta}$ is,  by definition, the $C^{*}-$ algebra generated by two unitaries symbols $X$ and $Y$ satisfying\cite{CGS}
\begin{equation}\label{e01}
	X\,Y = e^{2\pi i \theta}Y\,X.
\end{equation}
Let $\mathcal{H}$ be the Hilbert space   of square integrable  functions  $L^2$,  and  $\mathbb{T}^1$ the circle. 

There are many important representations $\pi_1$  and $\pi_2$ of the $C^{*}-$ algebra $\mathcal{D}_{\theta}$, given by the following relations\cite{CGS,S}:
\begin{itemize}
	\item Let $\mathcal{H} = L^2(\mathbb{T}^1)$,  $g\in\mathcal{H}$, $\lambda\in\mathbb{C}$,  and $v\in\mathbb{T}^1,$  then
	\begin{equation}\label{e02}
		(\pi_1(X)g)(v)=g(\lambda v) \quad \mbox{,}\quad 
		(\pi_1(Y)g)(v)= v\,g(v).
	\end{equation}
	\item Let $\mathcal{H} = L^2(\mathbb{T}^1)$,  $g\in\mathcal{H}$,  and $v\in\mathbb{T}^1,$ then
	\begin{equation}\label{e03}
		(\pi_2(Y)g)(v)=g(\bar{\lambda} v)\quad \mbox{,}\quad 
		(\pi_2(X)g)(v)= v\,g(v).
	\end{equation}
	\item Let $\mathcal{H} = L^2(\mathbb{R})$,  $g\in\mathcal{H}$, and $u\in\mathbb{R}.$ then
	\begin{equation}\label{e04}
		(\pi_3(X)g)(u)=g(u + 1) \quad \mbox{,}\quad 
		(\pi_3(Y)g)(u)= \lambda^u\,g(u).
	\end{equation}
\end{itemize}
There is a continuous action of $\mathbb{T}^2$, $\mathbb{T} = \mathbb{R}/2\pi\mathbb{Z}$, on $\mathcal{D}_{\theta}$ by $\mathcal{C}^*-$ algebra automorphism $\{\alpha_x\}$, $x\in\mathbb{R}^2$, defined as follows \cite{FGK}
\begin{equation}
	\alpha_x\big(X^{n_1}\,Y^{n_2}\big) = e^{ix\langle\,n_1 , n_2\, \rangle}X^{n_1}\,Y^{n_2},
\end{equation}
{where $n_1,n_2 \in\mathbb{Z}$}. 

The space of smooth elements for this action, that is those elements $a\in \mathcal{D}_{\theta}$ for which the map $x\mapsto \alpha_x(a)$ is $C^{\infty},$ will be denoted by $\mathcal{D}_{\theta}^{\infty}:\equiv C^{\infty}(\mathbb{T}^{2}_{\theta})$. It is a dense $\star$-subalgebra of $\mathcal{D}_{\theta}$ and can  alternatively be described as the space of elements of the  form $\displaystyle\sum_{n_1,n_2\in\mathbb{Z}}a_{n_1n_2}X^{n_1}\,Y^{n_2}$, {$a_{n_1n_2}\in\mathbb{R}^{*}$} with rapidly decreasing coefficients \cite{FK1}:
\begin{equation}
	\mathcal{D}_{\theta}^{\infty}\equiv \left\lbrace \sum_{n_1,n_2 \in \mathbb{Z}}a_{n_1n_2}X^{n_1}\,Y^{n_2};~~ \sup_{m,n \in \mathbb{Z}}(|n_1|^k |n_2|^q|a_{n_1n_2}|) < \infty, \,\, \forall k, q \in  \mathbb{Z}\right\rbrace .
\end{equation}
There exists a unique faithful trace $\varphi$ on $\mathcal{D}_{\theta}$ defined as follows \begin{equation}\label{p1}
	\varphi\left( \sum_{n_1,n_2 \in \mathbb{Z}}a_{n_1n_2}X^{n_1}\,Y^{n_2}\right) = a_{00}.
\end{equation}
We consider the inner product
\begin{equation}
	\langle \,u , v\, \rangle = \varphi(v^*u)\quad\mbox{,}\quad u,v \in\mathcal{D}_{\theta}.
\end{equation}
The derivations $d_j :\mathcal{D}_{\theta}^{\infty}\longrightarrow \mathcal{D}_{\theta}^{\infty}\mbox{,}\quad j\in\{1,2 \} $  associated to the group of automorphisms $\{ \alpha_x \}$ are given
by  the following relations \cite{CGS},
\begin{equation}
	d_1 (X) =   \, X \, , \quad d_1 (Y) = 0 \, ,
\end{equation}
\begin{equation}
	d_2 (X) = 0 \, , \quad d_2 (Y) =   \, Y \, .
\end{equation}

A theorem by Bratteli, Elliot, and Jorgensen \cite{BEJ} describes all the derivations of $\mathcal{D}_{\theta} $ which map $\mathcal{D}_{\theta}^{\infty}$ to itself: for almost all $\theta$ (Lebesgue), a derivation is of the form 
\begin{equation}
	d =\displaystyle\sum_{j=1}^{2}b_j d_j + [r,.],\quad \mbox{with}\quad r \in \mathcal{D}_{\theta}^{\infty},~b_j \in \mathbb{C}.
\end{equation}
Consider an orthonormal basis $e_1, e_2$ for $V$. Then the restriction of the above map $d$ defines commuting derivations
$d_i := d(e_i)  : \mathcal{D}_{\theta}^{\infty} \to \mathcal{D}_{\theta}^{\infty},  i\in\{1,2\} $  satisfying
\begin{equation}
	d_i(X_j)= d_{ij}X_i, \qquad  (i,j)\in \{1,2\}.
\end{equation}
The  $d_{j}$'s are the analogues of the differential operators $\frac{1}{i}\frac{\partial}{\partial u_{j}}$ acting on smooth functions on the ordinary torus. We have 
\begin{eqnarray}\label{ab}
	d_{j}(a^*) = -d_{j}(a)^*,
\end{eqnarray}
for $j\in\{ 1, 2 \}$ and $a\in\mathcal{D}_{\theta} ^{\infty}$. Moreover, since $\varphi \circ d_j =0$,
for all  $j$, we have the analogue of the formula for integration by parts:
\begin{equation}
	\varphi(ad_j(b)) = -\varphi(d_j(a)b),~~ a,b \in \mathcal{D}_{\theta}^{\infty}.
\end{equation}
Using all these derivations, we can define the Laplacian $\triangle : \mathcal{D}_{\theta}^{\infty} \to \mathcal{D}_{\theta}^{\infty}$ such that
\begin{equation}
	\triangle\equiv \sum_{i=1}^2 d_i^2.
\end{equation}
We note that the Laplacian $\triangle$ is independent  of
the orthonormal basis's choice $(e_1, e_2)$. All the above derivations are studied in the Hilbert space
 $\mathcal{H} = \textit{L}^2(\mathcal{D}_{\theta}, \varphi)$
 and the family $\{V^l\}_{l\in\mathbb{Z}^n}$ forms an orthonormal basis in $\mathcal{H}$\cite{N,S}. 
\begin{theorem}\cite{CGS}
	The canonical derivations
	$d_j$, $j\in\{1,2\}$ are self-adjoint on their  domains:
	\begin{equation}
		Dom(d_1)= \left\lbrace
		\sum a_{n_1n_2} X^{n_1}\,Y^{n_2} \mid \sum (1+  n_1^2 ) | a_{n_1n_2} |^2 < \infty \right\rbrace
	\end{equation}
	\begin{equation}
		Dom(d_2)= \left\lbrace \sum a_{n_1n_2} X^{n_1}\,Y^{n_2} \mid \sum (1+ n_2^2 ) | a_{n_1n_2}|^2 < \infty \right\rbrace,
	\end{equation}
	{ where $Dom(d)$ is the domain of $d$.}
\end{theorem} 
Moreover if we denote by $d_r := [r,.] $ with $r
\in\mathcal{D}_{\theta} \subset L^{\infty} ( \mathcal{D}_{\theta} , \varphi ) $ acting as left
multiplication in $\mathcal{H}$, then $ {d_r}^*=d_{r^*}\in\mathcal{B}(\mathcal{H}), $ { the space of bounded functions on $\mathcal{H}.$}
\section{ Noncommutative magnetic Laplacian and diffusion on $\mathcal{D}_{\theta}$}
\subsection{Construction of the magnetic quantum stochastic diffusion}
Let us now derive the quantum stochastic differential equation on  the $C^*-$ algebra $\mathcal{D}_{\theta}$ following  the  canonical construction of quantum stochastic flow or diffusion on a von Neumann or a $C^*-$ algebra $\mathcal{A}$ associated with a completely positive semigroup on $\mathcal{A}.$  For more details on such a canonical construction, see \cite{GS},  \cite{GSK} and references therein. Sauvageot in \cite {Sav} studied semigroups which have a local generator $\mathcal{L}$  characterized by:\begin{itemize}\label{df}
	\item $\mathcal{D} \subseteq Dom ( \mathcal{L} ) \subseteq \mathcal{A} \subseteq \mathcal{B} (
	\mathcal{H}) $, dense in $\mathcal{A} $ such that  $\mathcal{D} $ itself is a $*$-algebra;
	\item A $*$-representation $\pi$ in some Hilbert space $\mathcal{H}_1 $ and an
	associated $\pi $ derivation $\delta$ such that $\delta(x)\in\mathcal{B}
	(\mathcal{H},\mathcal{H}_1)$ and $\delta(xy)= \delta(x)y + \pi(x )\delta(y);$
	\item A second order cocycle relation: $ \mathcal{L}( x^*y) - {\mathcal{L}( x)}^*y -x^* \mathcal{L}(y) = {\delta(x)}^*\delta(y)$, for $ x,y\in\mathcal{D}$.
\end{itemize}
By analogy to the classical case, $\mathcal{L}$ is called the noncommutative Laplacian or Lindbladian. 

In this work concerning the magnetic field in the noncommutative $2-$torus, we consider the notation $d=\frac{\partial}{\partial t}$, the coordinate $(t_1,t_2)$ and $d_j=\frac{\partial}{\partial t_j}$, for each $j\in\{1,2\}$. 
We denote by $U(1)$ the unitary symmetric group.
\begin{definition}
	The component of the $U(1)$ gauge field is defined as follows:
	\begin{equation}\label{d1}
		G_k = \frac{1}{2}\sum_{j=1}^{2}\psi_{jk}t_j + \beta_k,
	\end{equation}
	where  $k\in\{1,2\},$ $\{\beta_k\}$ are real numbers and $\{\psi_{jk}\}$ are integers such that $\{\psi_{jk}=-\psi_{kj}\}$.
\end{definition}
The gauge $G=\displaystyle\sum_{k=1}^{2}G_kdt_k$ generates a uniform magnetic field
\begin{equation}
	B=\frac{1}{2}\sum_{j=1}^{2}\sum_{k=1}^{2}\psi_{jk}dt_j \wedge dt_k,
\end{equation}
\begin{definition}
	The magnetic Laplacian in a noncommutative $2-$ torus,  $\Delta^m : \mathcal{D}_{\theta}^{\infty} \longrightarrow \mathcal{D}_{\theta}^{\infty}$ is, defined by:
	\begin{equation}
		\Delta^m  := \sum_{j,k=1}^{2}g^{jk}\left(d_j-2\pi i G_j \right) \left(d_k-2\pi i G_k \right),
	\end{equation}
	where $g^{jk}$ is a metric.
\end{definition}
We denote by $\mathcal{L}^m$ the noncommutative magnetic Laplacian.
\begin{definition}\label{wa}
	The unperturbed  and   perturbed noncommutative magnetic Laplacians in a  noncommutative $2-$torus  are given, respectively, by:
	\begin{equation}\label{ul}
		\mathcal{L}^m_0 : = \sum_{j,k=1}^{2}g^{jk}\left(d_j-2\pi i G_j \right) \left(d_k-2\pi i G_k \right),
	\end{equation}
	and
	\begin{equation}\label{pl}
		\mathcal{L}^m := \sum_{j,k=1}^{2}g^{jk}\left(\delta_j-2\pi i G_j \right) \left(\delta_k-2\pi i G_k \right),
	\end{equation}
	where  $\delta_j = d_j  + d_{r_j}$ with $j\in\{1,2 \}$; $r\in\mathcal{D}_{\theta}$ and $d_j$ is the canonical derivation.
\end{definition}
The following Lemma gives the relation between the noncommutative magnetic Laplacian $(NCML),$ $\mathcal{L}^m_0$ (resp.  $\mathcal{L}^m$), and the ordinary noncommutative Laplacian $(NCL),$ $ \mathcal{L}_0$ (resp. $ \mathcal{L}.$)
\begin{lemma}\label{l1}
	The unperturbed  and   perturbed noncommutative magnetic Laplacians are given, respectively, by:
	\begin{itemize}
		\item 
		\begin{equation}\label{e13}
			\mathcal{L}^m_0 = \mathcal{L}_0 + T^0_{jk},
		\end{equation}
		where
		\begin{equation}\label{e14}
			T^0_{jk} =  \sum_{j=1}^{2}\left(\pi i G_j  d_j + 2\pi^2 G^2_j \right)
			+  \sum_{j\neq k}g^{jk}\left(d_j-2\pi i G_j \right) \left(d_k-2\pi i G_k \right),
		\end{equation}
		and
		\item	\begin{equation}\label{e15}
			\mathcal{L}^m = \mathcal{L} + T_{jk},
		\end{equation}
		with 
		\begin{equation}\label{e16}
			T_{jk} = \sum_{j=1}^{2}\left( \pi i G_j \delta_j + 2\pi^2 G^2_j \right)
			+  \sum_{j\neq k}g^{jk}\left(\delta_j-2\pi i G_j \right) \left(\delta_k-2\pi i G_k \right).
		\end{equation}
	\end{itemize}
\end{lemma}
\textit{Proof}:
\begin{itemize}
	\item Let $f$ be the test function. Then, according to the relation (\ref{ul}), and taking the metric $g^{jj}=-1/2$, we obtain 
	\begin{eqnarray}
		\sum_{j=1}^{2}g^{jj}(d_j-2\pi i G_j)^2f 
		&=& \mathcal{L}_0 f + \sum_{j=1}^{2}\Big(\pi i G_jd_j +2\pi^2G^2_j \Big)f.
	\end{eqnarray}
	Thus,
	{\begin{equation}
		\mathcal{L}^m_0 f = \mathcal{L}_0 f + \sum_{j=1}^{2}\Big(\pi i G_jd_j +2\pi^2G^2_j \Big)f+ \sum_{j\neq k}g^{jk}(d_j-2\pi i G_j)(d_k - 2\pi i G_k)f,
	\end{equation}}
	yielding  the relation (\ref{e14}).
	\item By Definition \ref{pl}, we get
	
	{\begin{equation}\label{e120}
		\sum_{j=1}^{2}g^{jj}(\delta_j-2\pi i G_j)^2f
		= \sum_{j=1}^{2}g^{jj}\Big(\delta_j(\delta_jf)  - \delta_j(2\pi i G_jf)- 2\pi i G_j\delta_jf - 4\pi^2G^2_jf\Big).
	\end{equation}}
	The computation of the terms in the right hand side of the above equation gives
	\begin{equation}\label{e125}
		\sum_{j=1}^{2}g^{jj}(\delta_j-2\pi i G_j)^2f
		= \sum_{j=1}^{2}g^{jj}\Big(\delta^2_jf - 2\pi i G_j\delta_jf - 4\pi^2G^2_jf\Big)
	\end{equation} 	
and	 the result holds.
	$\cqfd$
\end{itemize}

The next Proposition gives the construction of the unperturbed and perturbed magnetic quantum processes.
\begin{proposition}
	Let $\mathcal{D}_{\theta}^\infty$ be a $*-$ subalgebra of $\mathcal{D}_{\theta}.$  Consider a Hilbert space $\mathcal{H}$  and   a $*-$representation $\pi$ in $\mathcal{H}_1=\mathcal{H} \otimes \mathbb{C}^2 \cong \mathcal{H} \oplus \mathcal{H}$ defined by $\pi(y)=y\otimes I$. Then, the unperturbed and perturbed  magnetic quantum processes are driven by the triples  $(\pi, \delta^m_0, \mathcal{L}^m_0)$ and  $(\pi, \delta^m, \mathcal{L}^m),$ respectively.
\end{proposition}
\textit{Proof}:
\begin{enumerate}
	\item[(i)] Unperturbed magnetic quantum diffusion
	\begin{itemize}
		\item It is obvious to have $\mathcal{D}_{\theta}^\infty \subseteq Dom (\mathcal{L}^m_0)\subseteq \mathcal{D}_{\theta}\subseteq\mathcal{B} (\mathcal{H})$.
		\item 
		Taking $\delta^m_0 = \displaystyle\sum_{j=1}^{2}d_j$, one proves that, for $u,v \in\mathcal{D}_{\theta}^\infty$,  
		$\delta^m_0(u)\in\mathcal{B}(\mathcal{H},\mathcal{H}_1)$ and  
		\begin{equation}\label{e136}
			\delta^m_0(uv)= \delta^m_0(u)v + \pi(u)\delta^m_0(v).
		\end{equation}
		\item Moreover,
		using Lemma \ref{l1} and according to \cite{CGS}, we get  
		\begin{equation}
			\mathcal{L}^m_0(u^*v) = u^*\mathcal{L}_0(v) + \mathcal{L}_0(u)^*v + \delta_0(u)^*\delta_0(v) + T^0_{jk}(u^*v).
		\end{equation}
		It suffices to verify  $T^0_{jk}(u^*v)$.
		Using the same idea as in the above part, we get 		
		\begin{eqnarray}\label{jk}
			\sum_{j\neq k}g^{jk}\left(d_j-2\pi i G_j \right) \left(d_k-2\pi i G_k \right) &=& \sum_{j\neq k}g^{jk}\Big( d_jd_k - \Gamma_{jk}\nonumber\\ 
			&-& 2\pi i G_jd_k - 4\pi^2G_jG_k\Big).
		\end{eqnarray}
			with $\Gamma_{jk}f = \pi i \displaystyle\sum_{j=1}^{n}\psi_{jk}f $.
		Setting $C_1=4\pi^2G^2_j$ and $C_2=4\pi^2G_jG_k + \Gamma_{jk}$ and
		using the assumption that  $C_1$ and  $C_2$  satisfy the  Leibniz rule, we obtain 
		\begin{equation}\label{e141}
			T^0_{jk}(u^*v)= T^0_{jk}(u^*)v + u^*T^0_{jk}(v)
		\end{equation} 
and therefore, the result holds.
	\end{itemize}
	\item[(ii)]  Perturbed magnetic quantum diffusion
	
	For the fisrt relation, we perform the same computation as above and obtain:
	\begin{itemize}
		\item $\mathcal{D}_{\theta}^\infty \subseteq D (\mathcal{L}^m)\subseteq \mathcal{D}_{\theta}\subseteq\mathcal{B} (\mathcal{H})$.
		\item It is obvious to have
		$\delta^m(u)\in\mathcal{B}(\mathcal{H},\mathcal{H}_1)$ and
		\begin{equation}\label{e143}
			\delta^m(uv)= \delta^m(u)v + \pi(u)\delta^m(v),
		\end{equation} 
		with  $\delta^m = \displaystyle\sum_{j=1}^{2}(d_j + d_{r_j})$.
		
		\item
		
		According to the expression of the perturbed magnetic Laplacian $\mathcal{L}^m$, the Lemma \ref{l1}, and \cite{CGS},  we have 
		 
		\begin{equation}
			\mathcal{L}^m(u^*v)= u^*\mathcal{L}(v) + \mathcal{L}(u)^*v + \delta(u)^*\delta(v) + T_{jk}(u^*v)\quad\mbox{for}\quad (u,v) \in\mathcal{D}_{\theta}^\infty.
		\end{equation}
		Besides,
		\begin{eqnarray}
			T_{jk}(u^*v) &=& \sum_{j=1}^{2}\left( \pi i G_j \delta_j + 2\pi^2 G^2_j \right)(u^*v)\nonumber\\
			&+&  \sum_{j\neq k}g^{jk}\left(\delta_j-2\pi i G_j \right) \left(\delta_k-2\pi i G_k \right)(u^*v).
		\end{eqnarray}
		As $d_j$ is a derivation and $d_r =[r,.]$, $r\in\mathcal{D}_{\theta}$,  we obtain the same result as in equation (\ref{jk}).
		Thus, considering the assumption as in the case of $\mathcal{L}^m_0$, 
		and after some algebra, we obtain
		\begin{equation}\label{e146}
			T_{jk}(u^*v)=T_{jk}(u^*)v + u^*T_{jk}(v).
		\end{equation}
		Therefore, the result holds.
	\end{itemize}
	$\cqfd$
\end{enumerate}
Both the unperturbed and perturbed triples $(\pi,\delta^m_0, \mathcal{L}^m_0)$ and $(\pi,\delta^m,\mathcal{L}^m)$  satisfy previous properties $(1), (2), (3)$ giving rise to two quantum stochastic processes as stated  by the theorem established in  the following section.
\subsection{Magnetic quantum stochastic differential equation}
In this section, we derive the magnetic quantum stochastic differential equation  arising from the unperturbed and perturbed magnetic quantum stochastic processes, respectively.  More details on the quantum stochastic process can be found in \cite{E,EH,HR,PK} . {Let us first recall the following important statement used in the sequel:
	\begin{theorem}\cite{CGS}
		\begin{enumerate}
			\item The quantum stochastic differential equation for $y\in\mathcal{D}_{\theta}^{\infty }$
			\begin{equation}\label{e20}
			\left \{
			\begin{array}{l}
			\displaystyle
			d j_t^0 (y) = \sum_{k=1}^{2}j_t^0 ( i d_k (y) ) dw_k(t) +
			j_t^0 ( {\mathcal{L} }_0 (y))  dt, \\
			j_0^0 (y) = y \otimes I,\\
			\end{array}
			\right .
			\end{equation}
			has unique solution $j_t^0 $ which is a $*$-homomorphism from $\mathcal{D}_{\theta} $ 
			to 
			$
\mathcal{D}_{\theta} \otimes \mathcal{B} ( \Gamma ( L^2 ( R_+ ) \otimes C^2 )).$ Furthermore
			$j_t^0 (y) = \alpha_{( exp2 \pi i w_1 (t),exp2 \pi i w_2 (t))} (y),$
			where $( w_1,w_2 ) (t ) $ is the classical Brownian
			motion in dimension two and   $ E j_t^0 (y) = e^{t \mathcal{L}_0 } (y) $, where $E$ is the
			vacuum expectation in the
			Fock space $ \Gamma  ( L^2 ( \mathbb{R}_+ ) \otimes  C^2  ) $.  
			\item The quantum stochastic differential equation in $ \mathcal{H} \otimes \Gamma $:
			\begin{equation}\label{e20a}
			\left \{
			\begin{array}{l}
			\displaystyle
			d U_t = \sum_{k=1}^2 U_t\{ i {j_t}^0 (r_k ) d {A_k}^\dagger +
			i {j_t}^0 ({r_k}^*  )  d {A_k}  - \frac {1} {2} {j_t}^0 ( {r_k}^* r_k 
			)  dt  \}, \\
			U_0=I,\\
			\end{array}
			\right .
			\end{equation}
			has a unique unitary solution  \cite {EH}. Putting $ 
			j_t
			(y) = U_t j_t^0 (y) U_t^*, $ we obtain the quantum stochastic differential equation :
			\begin{equation}\label{e20b}
			d j_t (y) = \sum_{k=1 }^2 \{ j_t ( i {\delta }_k (y) ) d {A_k}^{\dagger
			} + j_t ( i {{\delta }_k}^{\dagger} (y) ) d {A_k} \} + j_t( \mathcal{L} (y))
			\end{equation}
			 and $ E j_t (y) = e^{t\mathcal{L} } (y).$
		\end{enumerate}
	\end{theorem}}

Then  our  main result in this section can be formulated as follows:
\begin{theorem}
	\begin{enumerate}
		\item[(i)] Let $j^0_t(x)$  be the solution of the quantum stochastic differential equation   {(\ref{e20})}. Then,  the  magnetic quantum stochastic differential equation for $x\in \mathcal{D}_{\theta}^{\infty}$
		\begin{equation}\label{e21}
			\left \{
			\begin{array}{l}
				\displaystyle
				df^{0}_{t}(x)=dj^0_t(x) + f^{0}_{t}(T^0_{jk}(x))dt,\quad (j,k)\in\{1,2\}  \\
				f^{0}_{0}(x)=x\otimes I,\\
			\end{array}
			\right .
		\end{equation}
		admits a unique solution $f^{0}_{t}$ which is a $*-$homomorphism defined as follows:
		$$f^{0}_{t}: \mathcal{D}_{\theta} \longrightarrow \mathcal{D}_{\theta}\otimes\mathcal{B}\left(\Gamma(L^{2}(\mathbb{R}_{+}))\otimes\mathbb{C}^2) \right)$$
		\begin{equation}\label{e22}
			f_t^0 (x) = j^0_t (x) + F^0_t(T^0_{jk}(x)),
		\end{equation}
		where $F^0_t$ is a primitive of $f^0_t$.
		
		Moreover,  the vacuum expectation in the Fock space $\Gamma(L^{2}(\mathbb{R}_+))\otimes\mathbb{C}^2)$ is given by:
		\begin{equation}\label{e23}
			Ef_t^0 (x) = \exp(t\mathcal{L}_0)(x) + F^0_t(T^0_{jk}(x)),
\, (j,k)\in\{1,2\}.
		\end{equation}
		\item[(ii)] Let $U_t$ be the solution of the quantum stochastic differential equation{(\ref{e20a})}. 
Then, the  magnetic quantum stochastic differential equation in $\mathcal{H}\otimes\Gamma$:
		\begin{equation}\label{e24}
			\left \{
			\begin{array}{l}
				\displaystyle
				dV_{t}= dU_t + V_{t}f^0_t(R_{lk})dt\\
				V_{0}=\mathbf{I},\\
			\end{array}
			\right .
		\end{equation}
		where
		\begin{eqnarray}\label{e25}
			R_{lk}= - 1/2\sum_{l=1}^{2}( r^*_l(2\pi i G_l) + 2\pi i G_lr_l)
			+ \sum_{l\neq k}(r^*_l - 2\pi i G_l)(r_k-2\pi i G_k), 
		\end{eqnarray}
		has a unique unitary solution.
		
		{Putting} $f_t(x)=V_tf^0_t(x)V^*_t$ leads to the following magnetic quantum stochastic differential equation:
		\begin{equation}\label{e26}
			df_t(x)=dg_t(x) +  f_t(\tilde{R}_{lk}(x))dt,
\quad (l,k)\in\{1,2\},
		\end{equation}
		where $\tilde{R}_{lk}(x) = R_{lk}x + xR^*_{lk},$  and $g_t(x)$ satisfies the quantum stochastic differential equation.
		
		Furthermore, the solution of (\ref{e26}) and the vacuum expectation are given by:
		\begin{equation}\label{e27}
			f_{t}(x) = j_{t}(x) + F_t(\tilde{R}_{lk}(x)),
		\end{equation}
		and
		\begin{equation}\label{e28}
			Ef_{t}(x)=\exp(t\mathcal{L})(x) + F_t(\tilde{R}_{lk}(x)),
		\end{equation}
respectively; $F_t$ is a primitive of $f_t$.
	\end{enumerate}
\end{theorem}
{\it Proof:}
\begin{enumerate}
	\item[(i)] From the construction of the unperturbed magnetic quantum process, we get
	\begin{equation}\label{e29}
		\left \{
		\begin{array}{l}
			\displaystyle
			df^{0}_{t}(x)=\sum_{k=1}^{2} f^{0}_{t}(id_k(x))dw_k(t) + f^{0}_{t}(\mathcal{L}^m_{0}(x))dt  \\
			f^{0}_{0}(x)=x\otimes I,\\
		\end{array}
		\right .
	\end{equation}
	where $(w_k)_{k\in\{1,2\}}$ is a classical $2-$ dimensional Brownian motion.
	
	{Since the noncommutative Brownian motion is   $dw= dA + dA^{\dagger},$ thus} the equation
	(\ref{e29}) is equivalent to 
	\begin{equation}\label{s3}
		\left \{
		\begin{array}{l}
			df^{0}_{t}(x)=f^0_{t}(\delta^{{\dagger}}(x))dA + f^0_{t}(\delta(x))dA^{{\dagger}} + f^{0}_{t}(\eta(x))dt,  \\
			f^{0}_{0}(x)=x\otimes I\\
		\end{array}
		\right .
	\end{equation}
	where $\delta$ and $\eta$ are the map structures satisfying 
	\begin{equation}
		\eta(xy)=\eta(x)y + x\eta(y) + \delta^{{\dagger}}(x)\delta(y).
	\end{equation}
	Moreover,  if $\delta$ and $\eta$ are bounded in $\mathcal{B}(\mathcal{H})$, then there exists a unique quantum diffusion $f^0(t)$ which satisfies (\ref{s3})\cite{EH}. In our case, 
	\begin{equation}
		\delta(x)=\sum_{k=1}^2 id_k(x)\quad\mbox{and}\quad  \eta(x)=\mathcal{L}^m_0(x),
	\end{equation}
	both satisfying 
	\begin{equation}
		\mathcal{L}^m_0(x^*y) - \mathcal{L}^m_0(x^*)y - x^*\mathcal{L}^m_0(y)= \delta(x)^*\delta(y).
	\end{equation}
	since  $d_j :\mathcal{D}_{\theta}^\infty \longrightarrow \mathcal{D}_{\theta}^\infty $, $j\in\{1,2\}$ are the canonical derivations. By definition of $\mathcal{L}^m_0$, we deduce that both the two operators are bounded.
	Therefore, the equation (\ref{s1}) admits a unique quantum diffusion $f_t^0$.
	
	According to the Lemma (\ref{l1}), the equation (\ref{e29}) can be also written as follows:  
	\begin{equation}\label{e210}
		\left \{
		\begin{array}{l}
			\displaystyle
			df^{0}_{t}(x)=\sum_{k=1}^{2} f^{0}_{t}(id_k(x))dw_k(t) + f^{0}_{t}(\mathcal{L}_{0}(x))dt + f^{0}_{t}(T^0_{jk}(x))dt  \\
			f^{0}_{0}(x)=x\otimes I.\\
		\end{array}
		\right .
	\end{equation}
	According to \cite{CGS}, the equation (\ref{e21}) holds,
	since $dg^0_t(x)$ admits a unique solution \cite{CGS}. Therefore, the equation (\ref{e21}) has a unique solution.
	
	Using the relation (\ref{e29}) and the linearity of the vacuum expectation, we obtain 
	\begin{equation}\label{eh}
		Ef^{0}_{t}(x) = f^{0}_{0}(x) + \int_{0}^{t}E\left(h^{0}_{s}(\mathcal{L}^m_0(x))\right)ds.
	\end{equation}
	A similar computation as in \cite{PK} implies that
	\begin{eqnarray}
		Ef^0_{t}(x) 
		&=& e^{t\mathcal{L}^m_0}(x).
	\end{eqnarray}
	
	Besides, as $f^{0}_{0}(x)= g^{0}_{0}(x)$, the integral form of (\ref{e21}) is given by:
	\begin{eqnarray}\label{e212}
		f^{0}_{t}(x) &=& j^{0}_{t}(x) +  \int_{0}^{t}f^0_s(T^0_{jk}(x))ds, 
	\end{eqnarray}	
	and considering the primitive $F^0_t$ of $f^{0}_{t}$, we have (\ref{e22}).
	Using the linearity of the vacuum expectation, we obtain
	\begin{equation}\label{e213}
		Ef^{0}_{t}(x) = Ej^{0}_{t}(x) +  EF^0_t(T^0_{jk}(x)),
	\end{equation}
	and, according  to \cite{CGS}, we obtain
	the result.
	\item[(ii)] By definition of the quantum Brownion motion and setting  $d_l = r_l$, for $l\in\{1,2\},$ 
	we obtain, from the relation (\ref{e29}), the following equation:
	\begin{equation}\label{e215}
		\left \{
		\begin{array}{l}
			\displaystyle
			dV_{t}=\sum_{l=1}^{2}V_{t}\Big(if^{0}_{t}(r_{l})dA^{{\dagger}}_{l}+i f^{0}_{t}(r^{*}_{l})dA_{l} + f^0_t(\mathcal{L}^m_0)dt \Big)\\
			V_{0}=\mathbf{I},\\
		\end{array}
		\right .
	\end{equation}
	
	Setting 
	\begin{equation}
		dV^k_t=V_t\left\lbrace i h^0_t(r_k)dA^{{\dagger}}_k+i h^0_t(r^*_k)dA_k+ f^0_t(\mathcal{L}^m_0)dt \right\rbrace,
	\end{equation}
	where $k\in\{1,2\},$
	then 
	\begin{equation}
		dV_t=dV^1_t+ dV^2_t. 
	\end{equation}
	But,  the equation $dV^1_t$ is equivalent to
	\begin{equation}\label{q1}
		dV^1_t = V_tf^0_t(\lambda^1_\mu)d\Lambda^{\mu}_1,
	\end{equation} 
	and using \cite{EH}, we conclude that (\ref{q1}) admits a unique solution, and for $k\in\{1, 2 \}$ 
	\begin{equation}
		dV^k_t = V_th^0_t(\lambda^k_\mu)d\Lambda^{\mu}_k
	\end{equation}
	admits a unique solution. Therefore, (\ref{e215}) admits a unique solution. 
	
	Now, we show the unitarity of the solution. As $V_t$ is solution of the (\ref{e215}),  the adjoint $V^*_t$ is also solution of the equation (\ref{e215}),  and we have
	\begin{eqnarray}
		dV^*_t &=& \sum_{l=1}^2( -i f^{0}_t(r^*_l)dA_l - i f^0_t(r_l)dA^{{\dagger}}_l+ f^0_t(\mathcal{L}^m_0)V^*_t.
	\end{eqnarray} 
	Using It$\hat{o}$'s formula and table, we have $d(V_tV^*_t) =0$. 
	Thus, $V_tV^*_t= V_0$, but $V_0 =I$, and, therefore,  the result follows.
		
	Furthermore, taking the metric $g^{ll}=-1/2$, for $l\in\{1,2\}$ and after computation, we obtain
	\begin{eqnarray}
		dV_{t}
		&=& dU_t + V_tf^0_t(R_{lk})dt.
	\end{eqnarray}
	Thus, we obtain the relation  (\ref{e24}). Using \cite{CGS}, the result holds.
	
	Putting $f_t(x)=V_tf^0_t(x)V^*_t$, and according to  It$\hat{o}$'s formula and table \cite{EH} 
	\begin{eqnarray}
		df_t(x)  
		&=& \sum_{l=1}^2 f_t\left( i[r_l , x]\right)dA^{{\dagger}}_l + \sum_{l=1}^2 f_t\left( i[r^*_l , x]\right)dA_l
		+ \sum_{l=1}^2 f_t(\mathcal{L}^m(x))dt,
	\end{eqnarray}
	with $\mathcal{L}^m(x) = xH_{lk} - H^*_{lk}x + xr^{*}_{l}r_{l}.$ 
	According to \cite{CGS}, we obtain the relation (\ref{e26})  
	
	Moreover, using the integral form and the linearity of the vacuum expectation, we get  (\ref{e27}) and (\ref{e28}), respectively.
\end{enumerate}
$\cqfd$

\section{Weyl asymptotics for $\mathcal{D}_{\theta}$} 

Let us define the volume and the integrated scalar curvature of $\mathcal{D}_{\theta}$ using the Weyl asymptotics, and  their invariance under the perturbation  from $\mathcal{L}^m_{0}$ to $\mathcal{L}^m$. 
For a general discussion on Weyl asymptotics for torus, see \cite{Sav}, \cite{S}, and \cite{Con} (and references therein).  

The magnetic expectation of the Brownian motion on the manifold is considered as the magnetic heat semigroup 
$\mathcal{T}^m_t$ with the magnetic Laplace-Beltrami operator $\Delta^m$ as its generator. According to \cite{Ros}, we consider that $\mathcal{T}^m_t$ is a magnetic integral operator on the space of square integrable functions  on $M$ to $dvol,$ $\big( L^2(M, dvol)\big),$ and has the magnetic integral kernel $\mathcal{T}^m_t(s,h)$. The volume of the manifold is defined as follows\cite{S}:
\begin{equation}
Vol (M):=\int_{M}\mathcal{T}^{m(0}(e,e)\,dvol(e).
\end{equation}
Using \cite{Ros}, the smooth integral kernel admits an asymptotic expansion as $t \to 0^+$.
Thus,
\begin{eqnarray}
Vol(M) &=& \int_{M}\Big(\mathcal{T}^m_t(e,e)t^{d/2} - \sum_{l=1}^{\infty}\mathcal{T}^{m(l)}(e,e)t^{-d/2 + l} \Big)dvol(e)\nonumber\\
&=& \lim_{t \to 0^+}t^{d/2}\int_{M}\mathcal{T}^m_t(e,e)dvol(e).
\end{eqnarray}
Taking the trace in the space $\big( L^2(M, dvol)\big)$, we obtain
\begin{equation}
Vol(M) = \lim_{t \to 0^+}t^{d/2} T_r \mathcal{T}^m_t.
\end{equation}
Moreover, the scalar curvature  at $e\in M$is given by $s(e)= 1/6\,\mathcal{T}^{m(1)}(e,e)$.

Thus, the integrated scalar curvature is defined as follows:
\begin{eqnarray}
s &=& \int_{M} s(e) dvol(e)\nonumber\\
&=& \frac{1}{6}\lim_{t \to 0^+}\,t^{d/2 -1}[T_r\,\mathcal{T}^m_t - t^{-d/2}Vol(M)].
\end{eqnarray}

In the classical case, the magnetic heat semigroup $T^m_t$ is replaced by the unperturbed magnetic expectation $e^{t\mathcal{L}^m_0}$ and the perturbed magnetic expectation $e^{t\mathcal{L}^m}$, respectively. According to Lemma \ref{l1} and the fact that $\mathcal{L}_0$ and $T^0_{jk}$ are two commuting operators, we have 
\begin{equation}
\mathcal{T}^m_t = e^{t\,T^0_{jk}}\mathcal{T}_t\quad\mbox{and}\quad \mathcal{T}^m_t = e^{t\,T_{jk}}\mathcal{T}_t.
\end{equation}

For the estimation of the  magnetic expectations, we need to study both the unperturbed and  perturbed magnetic Laplacians. In the sequel, we denote by $\mathcal{B}_p$ the Schatten ideals in $\mathcal{B}(\mathcal{H})$.

Before giving the Proposition summarizing the properties of the unperturbed and perturbed noncommutative magnetic operators in the case of the noncommutative $2-$ torus, let us recall the following important definition and theorem.
\begin{definition}\cite{RS}
	Let $T : Dom(T)\longrightarrow \mathcal{H}$ be a self-adjoint operator and let  $F : Dom(F)\longrightarrow \mathcal{H}$ be symmetric. We say that $F$ is $T-$ bounded with bound $\alpha >0$ if $Dom(T)\subset Dom(F)$ and there exists $\beta >0$ so that: 
	\begin{equation}\label{e33}
	\left\|F h \right\| \leq \alpha \left\|Th \right\| + \beta\left\|h \right\|,    
	\end{equation}
	for all $h\in Dom(T)$.
\end{definition}
\begin{theorem}[Kato-Rellich theorem]\cite{RS}
	Suppose that $T$ is self-adjoint, $F$ is symmetric, and $F$ is $T-$ bounded with relative bound $\alpha \leq 1$. Then, $T+F$ is self-adjoint on $Dom(T)$ and essentially self-adjoint on any core of $T$. Further, if $T$ is bounded below by $M$, then, $T+F$ is bounded below by $M-Max\{\frac{\beta}{1-\alpha}, \alpha \left| M\right| + \beta \}$.
\end{theorem}
\begin{proposition}\label{P1}
	\begin{enumerate}
		\item $\mathcal{L}^m_0$ is a self-adjoint operator in $L^2(\varphi)$ and for $n_1,n_2 \in\mathbb{Z}$, we have
		\begin{equation}\label{kr}
		\mathcal{L}^m_0(X^{n_1}\,Y^{n_2}) = \Big(\frac{1}{2}(n^2_1 + n^2_2) + \eta_{jk}(n_1,n_2)\Big)(X^{n_1}\,Y^{n_2}),
		\end{equation}
		where
		\begin{equation}
		\eta_{jk}(n_1,n_2)= \sum_{j=1}^{2}\left(\pi i G_jn_j + 2\pi^2 G^2_j \right)
		+  \sum_{j\neq k}g^{jk}\big(n_j-2\pi i G_j \big)\big(n_k-2\pi i G_k \big).
		\end{equation}
		\item For $r_j\in\mathcal{D}^{\infty}_{\theta}$, $j\in\{ 1,2\},$ and self-adjoint, then 
		
		(i)	\begin{equation}
		\mathcal{L}^m = \mathcal{L}^m_0 + T^1_{jk} + T^2_{jk},
		\end{equation}
		with
		\begin{equation}\label{e18}
		T^1_{jk}=\displaystyle
		\sum_{j=1}^{2}g^{jj}\Big(d^2_{r_j} - 2\pi i G_jd_{r_j} \Big)
		-\sum_{j\neq k}g^{jk}\Big(2\pi i G_kd_{r_j} 
		+ 2\pi i G_jd_{r_k}\Big)
		\end{equation}
		or
		\begin{equation}\label{e18a}
		T^1_{jk}=	\displaystyle\sum_{j=1}^{2}g^{jj}(d^2_{r_j} + d_{d_j(r_j)} - 2\pi i G_jd_{r_j} )
		-2\pi i\sum_{j\neq k}g^{jk}( G_kd_{r_j} 
		+  G_jd_{r_k}),
		\end{equation}
		and 
		\begin{equation}\label{e19}
		T^2_{jk}=
		\left \{
		\begin{array}{l}
		\displaystyle
		\sum_{j=1}^{2}g^{jj}(d_{r_j}d_j + d_jd_{r_j}  ) + \sum_{j\neq k}g^{jk}\left( d_jd_{r_k} +d_{r_j}d_k + d_{r_j}d_{r_k} \right)\\
		\mbox{or} \\
		\displaystyle 2\sum_{j=1}^{2}g^{jj}d_{r_j}d_j + \sum_{j\neq k}g^{jk}\left( d_jd_{r_k} + d_{r_j}d_k + d_{r_j}d_{r_k} \right).
		\end{array}
		\right .
		\end{equation}
		(ii) $T^2_{jk}$ is compact relative to $\mathcal{L}^m_0$.\\
		(iii) $\mathcal{L}^m$ is self-adjoint on $Dom \big( \mathcal{L}^m_0\big)$ and has a compact resolvent.
	\end{enumerate}
\end{proposition}
{\it Proof:}
\begin{enumerate}
	\item According to Lemma \ref{l1} and the fact that
	 $d_j$, $j\in\{1,2\}$ are self-adjoint on their domains, we deduce that for $(j,k)\in\{1,2\}$, $T^0_{jk}$ is self-adjoint on $\mathcal{D}^{\infty}_{\theta}\subseteq L^2(\varphi)$ and using \cite{CGS}, we conclude that $\mathcal{L}^m_0$ is self-adjoint. Besides, according to \cite{S}, 
	 we know the expression of $\mathcal{L}_0(X^{n_1}Y^{n_2})$ for $n_1,n_2 \in\mathbb{Z}$ and $(X,Y)\in\mathcal{D}_{\theta}$.
It is obvious to compute  $T^0_{jk}(X^{n_1}Y^{n_2})$ and the relation (\ref{kr}) holds.
	\item (i). Since $\delta = d + d_r$, and according to (\ref{pl}), 
	we obtain, respectively,
	\begin{eqnarray}\label{e129}
	\sum_{j=1}^{2}g^{jj}(\delta_j-2\pi i G_j)^2f &=& \sum_{j=1}^{2}g^{jj}\Big(d^2_j +d^2_{r_j}+ d_jd_{r_j} + d_{r_j}d_j\nonumber\\ &-& 2\pi i G_jd_j
	- 2\pi i G_jd_{r_j} - 4\pi^2G^2_j \Big)f
	\end{eqnarray}
and
	\begin{eqnarray}\label{e134}
	\sum_{j\neq k}g^{jk}(\delta_j-2\pi i G_j)(\delta_k - 2\pi i G_k)f &=&\sum_{j\neq k}g^{jk}\Big(d_jd_k + d_jd_{r_k}\nonumber\\ 
	&+& d_{r_j}d_k
	+ d_{r_j}d_{r_k}\nonumber\\ 
	&-& 2\pi i d_jG_k - 2\pi i G_jd_k\nonumber\\
	&-&4\pi^2G_jG_k 
	- 2\pi i d_{r_j}G_k\nonumber\\ &-& 2\pi i G_jd_{r_k} \Big)f.
	\end{eqnarray}
	Using the relations (\ref{e129}) and (\ref{e134}),   
	 the result holds.
	
	(ii) Taking $g^{11}= g^{22}=-1/2$, we get 
	\begin{equation}
	T^2_{jk}\big( -\mathcal{L}^m_0 + n^2\big) = A\big( -\mathcal{L}^m_0 + n^2\big) + K_{jk}\big( -\mathcal{L}^m_0 + n^2\big).
	\end{equation}
	According to Lemma \ref{l1}, we have
	\begin{eqnarray}
	A\big( -\mathcal{L}^m_0 + n^2\big) 
	&=& A\big( -\mathcal{L}_0 + N^2\big),
	\end{eqnarray}
	with $N^2 = n^2 - T^0_{jk}$. Using \cite{S}, $A\big( -\mathcal{L}_0 + N^2\big)$ is compact relative  to $\mathcal{L}_0$. As the operator $K_{jk}$ is similar to $A$, we deduce that $T^2_{jk}$ is compact relative to $\mathcal{L}^m_0$. 
	\item  Since $\mathcal{L}^m_0$ is self-adjoint in $L^2\big(\varphi \big)$ and 
	\begin{equation}\label{ac}
	(\mathcal{L}^m - \mathcal{L}^m_0)(n^2 - \mathcal{L}^m_0)^{-1} \to  0 \quad\mbox{in operator norm}\quad \mbox{as}\quad n \to \infty,
	\end{equation}	
	according to Lemma \ref{l1} and Kato-Rellich theorem, $\mathcal{L}^m$ is self-adjoint. Furthermore, 
	\begin{equation}
	\big(-\mathcal{L}^m + n^2 \big)^{-1} = \big(-\mathcal{L} + f(n) \big)^{-1},
	\end{equation}
	where $f(n) = n^2 - T^1_{jk}$. Using \cite{CGS}, $\mathcal{L}$ has a compact resolvent. Thus,  we deduce that $\mathcal{L}^m$ has a compact resolvent.
	
	Let us show that $\big( \mathcal{L}^m - n^2\big)^{-1} - \big( \mathcal{L}_0^m - n^2\big)^{-1} $ is a trace class.  For $u\in\rho(\mathcal{L}^m)\cap \rho(\mathcal{L}^m_0)$, we have
	\begin{equation}
	(\mathcal{L}^m - u)^{-1} - (\mathcal{L}_0^m - u)^{-1} =
	\big(\mathcal{L} - g(u)\big)^{-1} - \big( \mathcal{L}_0 - g_0(u)\big)^{-1},
	\end{equation}
	where $g(u) = u- T_{jk}$  and $g_0(u) = u - T^0_{jk}$.
	
	Since $u\in\rho(\mathcal{L})$, then $g(u)\in\rho(\mathcal{L})\subset\rho(\mathcal{L}^m)$. Similarly, we show that $g_0(u)\in\rho(\mathcal{L}_0^m)$. Thus, 
	
	\begin{eqnarray}
	(\mathcal{L}^m - u)^{-1} - ( \mathcal{L}_0^m - u)^{-1} &=& (\mathcal{L}_0^m - u)^{-1}\big[(\mathcal{L}^m - \mathcal{L}_0^m)\mathcal{L}_0^m - u)^{-1} \nonumber\\ &+& 1\big]^{-1}
	(\mathcal{L}^m_0 - \mathcal{L}^m)(\mathcal{L}^m_0)^{-1},
	\end{eqnarray}
	since $\big(\mathcal{L}^m - \mathcal{L}^m_0\big)\big(\mathcal{L}^m_0 + n^2\big)^{-\frac{1}{2}}$ is bounded. In fact,
	using the Proposition \ref{p1}, we have
	\begin{equation}
	\big(\mathcal{L}^m - \mathcal{L}^m_0\big)\big(\mathcal{L}^m_0 + n^2\big)^{-\frac{1}{2}} = \big(T^1_{jk} + T^2_{jk}\big)\big(\mathcal{L}^m_0 + n^2\big)^{-\frac{1}{2}},
	\end{equation}
	and taking the norm, we get the result. It follows that $\big( \mathcal{L}^m_0 + n^2\big)^{-\frac{1}{2}}\in\mathcal{B}_3\big( L^2(\varphi)\big).$ Moreover, $\big(\mathcal{L}^m_0 - u\big)^{-1}\in\mathcal{B}_3/2\big( L^2(\varphi)\big)$. 
	Therefore,  we deduce that
	$\big( \mathcal{L}^m - n^2\big)^{-1} - \big( \mathcal{L}_0^m - n^2\big)^{-1} $ is a trace class.
	$\cqfd$
\end{enumerate}

In  the next theorem, we investigate the  invariance of volume and integrated scalar curvature under the perturbation of $\mathcal{L}^m_0$ from $\mathcal{L}^m$. 
\begin{definition}
	The volume and integrated scalar curvature in noncommutative 2-torus are given, respectively,  by :
	\begin{equation}\label {v}
	V(\mathcal{D}_{\theta}):= \lim_{t \longrightarrow 0+ } tTr \mathcal{T}^m_t,
	\end{equation}
	\begin{equation}\label {is}
	s(\mathcal{D}_{\theta}):= \frac {1} {6} \lim_{t \longrightarrow 0+ } [ Tr \mathcal{T}^m_t -t^{-1} V ].
	\end{equation}
\end{definition}
\begin{theorem} Let $V$ and $s$ be the volume and integrated scalar curvature, respectively, on $\mathcal{D}_{\theta}.$ Then,
	\begin{enumerate}
		\item The volume V of $\mathcal{D}_{\theta} $ 
		is invariant under the perturbation from $\mathcal{L}^m_0$ to $\mathcal{L}^m$.
		\item The integrated scalar curvature is not invariant under the same perturbation.
	\end{enumerate}
\end{theorem}
\textit{Proof:}
\begin{enumerate}
	\item Using the relation (\ref{v}), we have
	  \begin{equation} \label {81}
	V(\mathcal{L}^m) -V(\mathcal{L}^m_0 ) = \lim_{t\rightarrow 0+}tTr(e^{t\mathcal{L}^m} - e^{t \mathcal{L}^m_0}).
	\end{equation}
	Then, we need to compute $Tr(e^{t \mathcal{L}^m} - e^{t \mathcal{L}^m_0 } )$. For that, we distinguish two cases  of $r_j$.
	
	(i) For $r_j\in\mathcal{D}^{\infty}_{\theta}$, $j\in\{1,2\}$ and $0<v\leq t \leq 1$,  we have
	\begin{equation}
	e^{t\mathcal{L}^m} - e^{t\mathcal{L}^m_0} =  \int_0^t e^{(t-v)\mathcal{L}^m} (\mathcal{L}^m -\mathcal{L}^m_0)e^{v \mathcal{L}^m_0}dv,
	\end{equation}
	and using $2$ iterations, we get
	\begin{eqnarray}\label{2i}
	e^{t\mathcal{L}^m} - e^{t \mathcal{L}^m_0}&=& \int_0^t e^{(t-v) \mathcal{L}^m_0 } (\mathcal{L}^m -\mathcal{L}^m_0) e^{v \mathcal{L}^m_0 } dv -\int_0^t dt_1 e^{(t-t_1 ) \mathcal{L}_0 } (\mathcal{L}^m-\mathcal{L}^m_0 )\nonumber\\
	&\times& \int_0^{t_1} dt_2 e^{(t_1 -t_2 ) \mathcal{L}^m_0 } ( \mathcal{L}^m - \mathcal{L}^m_0 ) e^{ t_2
		\mathcal{L}^m_0 }\nonumber\\
	&-&\int_0^t dt_1 e^{( t- t_1) \mathcal{L}^m } ( \mathcal{L}^m -\mathcal{L}^m_0 )\int_0^{t_1} dt_2 e^{ (t_1 -t_2 )\mathcal{L}^m_0 } (\mathcal{L}^m- \mathcal{L}^m_0 )\nonumber\\
	&\times&  \int_0^{t_2} dt_{3} e^{( t_2 -t_{3} ) \mathcal{L}^m_0 } ( \mathcal{L}^m - \mathcal{L}^m_0 ) e^{ t_{3} \mathcal{L}^m_0 }\nonumber\\
	&\equiv& J_1 (t) + J_2 (t)  + J_{3} (t).
	\end{eqnarray}
	Using the trace norm, we estimate all these terms. For that, we need to get the $\mathcal{B}_p-$norm of   $(\mathcal{L}^m-\mathcal{L}^m_0 ) e^{t \mathcal{L}^m_0 }$.
	For $0\leq v\leq 1$, we have
	\begin{eqnarray}\label{es}
	\left\| (\mathcal{L}^m - \mathcal{L}^m_0 )e^{ v\mathcal{L}^m_0 }\right\|_p &\leq& \left\|e^{ vT^0_{jk} }\right\|\, v^{-\frac{1}{p} - \frac{1}{2}}\quad\mbox{,}\quad (j,k)\in\{1,2\}.
	\end{eqnarray}
	We estimate the terms in (\ref{2i}), using the H$\ddot{o}$lder inequality for Schatten norms and the estimation of $\left\|I_1(t)\right\|_1$, $\left\|I_2(t)\right\|_1$,  and $\left\|I_3(t)\right\|_1$ in \cite{CGS}. We get,
	\begin{equation}
	\left\|J_1(t)\right\|_1 \leq  \int_{0}^{t}\left\| e^{(t-v)\mathcal{L}^m_0}\right\|_{p_1} \left\| ( \mathcal{L}^m -\mathcal{L}^m_0) e^{v \mathcal{L}^m_0 }\right\|_{p_2}dv.
	\end{equation}
	According to (\ref{es}), 
and \cite{CGS,S}, we get the following relation
	\begin{equation}
	\left\|J_1(t)\right\|_1
	\leq \alpha_1\,\beta_1\,t^{- 1/2},
	\end{equation}
		where  $\alpha_1 = \left\|e^{(t-v)T^0_{jk}} \right\|$ and $\beta_1 = \left\|e^{v\,T^0_{jk}} \right\|$.
	Similarly,
	\begin{equation}\label{j2}
	\left\|J_2(t)\right\|_1 \leq \alpha_2\,\beta_2\,\gamma_2\,c,
	\end{equation}
	where $c\in\mathbb{R}_{+}^{*}$ 
	, $\alpha_2 = \left\|e^{(t-t_1)T^0_{jk}} \right\|$, $\beta_2 = \left\|e^{(t_1-t_2)T^0_{jk}} \right\|$, and $\gamma_2 = \left\|e^{t_2T^0_{jk}} \right\|$.
	
	Considering the following relation, 
	\begin{equation}
	\left\|(\mathcal{L}^m - n^2)^{-1}\right\|\leq \frac{2}{n}\quad\mbox{,}\quad n\in\mathbb{N}^{*},
	\end{equation}
	we obtain the following estimation for $J_3(t)$
	\begin{equation}
	\left\|J_3(t)\right\|_1\leq \alpha_3\,\beta_2\,\gamma_3\left\|I_3(t)\right\|_1,
	\end{equation}
	where $\alpha_3 = \left\|e^{ (t_2-t_3)T^0_{jk} }\right\|$ and $\gamma_3 =\left\|e^{ t_3T^0_{jk} }\right\| $.
	
	Since, $\left\|I_3(t)\right\|_1\longrightarrow 0$ as $t\longrightarrow 0^+$ (see \cite{CGS}),
	then, we deduce that  $\left\|J_3(t)\right\|_1\longrightarrow 0$ as $t\longrightarrow 0^+$. 
	
	Therefore, 
\begin{equation}
	V( \mathcal{L}^m) = V(\mathcal{L}^m_0 ).
	\end{equation} 
	
	(ii) For $r_j\in\mathcal{D}_{\theta}$, we have 
	\begin{equation}\label{rd}
	\mathcal{L}^m - \mathcal{L}^m_0 = T^1_{jk} + \sum_{j=1}^{2}g^{jj}\big( d_jB_j + C_jd_j\big) +
	\sum_{j\neq k}g^{jk}\big(d_jB_k + C_jd_k + C_jB_k\big), 
	\end{equation}
	where $T^1_{jk}, B_j, C_j,$ and $B_k$ are bounded.
	
	Let us compute all the integrals in the equation (\ref{2i}).
	\begin{itemize}
		\item According to the expression of $J_1(t)$ and the relation (\ref{rd}), we have
		\begin{eqnarray}
		e^{(t-v)\mathcal{L}^m_0}(\mathcal{L}^m - \mathcal{L}^m_0)e^{v\mathcal{L}^m_0} &=& e^{(t-v)\mathcal{L}^m_0}( T^1_{jk} + K_1 + K_2) e^{v\mathcal{L}^m_0}\nonumber\\
		&=& E_1 + E_2 + E_3.
		\end{eqnarray} 
		Thus, 
		\begin{equation}
		J_1(t) = \int_{0}^{t} E_1\,dv + \int_{0}^{t} E_2\,dv + \int_{0}^{t} E_3\,dv .
		\end{equation}
		Besides,
		\begin{eqnarray}
		\int_{0}^{t} E_1\,dv  
		&=& T^1_{jk} \,I_1(t),
		\end{eqnarray}
		where $I_1(t)$ is given in \cite{CGS,S}.
		
		Thus,
		\begin{equation}
		\left\|\int_{0}^{t} E_1\,dv\right\|_1  \leq \left\| T^1_{jk}\right\|\left\|I_1(t)\right\|_1. 
		\end{equation}
		Moreover,
		\begin{eqnarray}
		\int_{0}^{t} E_2\,dv 
		&=& \sum_{j=1}^{2}g^{jj}\int_{0}^{t} e^{(t-v)\mathcal{L}^m_0}\, d_jB_j\,e^{v\mathcal{L}^m_0}\,dv
		\nonumber\\ 
		&+& \sum_{j=1}^{2}g^{jj}\int_{0}^{t} e^{(t-v)\mathcal{L}^m_0}\,C_jd_j\,e^{v\mathcal{L}^m_0}\,dv
		\end{eqnarray}
		and, after computation, we get
			\begin{eqnarray}
			\left\|\int_{0}^{t} E_2\,dv\right\|_1  &\leq& \mu\,\sum_{j=1}^{2}g^{jj}\alpha_j\big(\left\|B_j\right\|+\left\|C_j\right\|\big)\,t,
			\end{eqnarray}
				with $\mu\in\mathbb{R}_+^*.$ 
		
		Furthermore,
		\begin{eqnarray}
		\int_{0}^{t} E_3\,dv 
		&=& \int_{0}^{t}e^{(t-v)\mathcal{L}^m_0}\sum_{j\neq k}g^{jk}(d_jB_k + C_jd_k + C_jB_k)e^{v\mathcal{L}^m_0}dv.
		\end{eqnarray}
		Analogously to the above calculation, we obtain
		
		\begin{eqnarray}
		\left\|\int_{0}^{t} E_3\,dv \right\|_1  &\leq& \nu\,\sum_{j\neq k}g^{jk}\alpha_j\,\left\| B_k\right\| \,t + \alpha_k\,\left\| C_j\right\| \,t + \left\|C_jB_k\right\|, 
	\end{eqnarray}
	with $\nu\in\mathbb{R}^*_+.$

		Therefore,
		
		\begin{eqnarray}\label{J1}
		\left\|J_1(t)\right\|_1 &\leq& \mu\sum_{j=1}^{2}g^{jj}\alpha_j\Big(\left\|B_j\right\| + \left\|C_j\right\|\Big)\,t
		\nonumber\\
		&+& \nu\,\sum_{j\neq k}g^{jk}\Big(\big( \alpha_j\,\left\| B_k\right\|  + \alpha_k\,\left\| C_j\right\|\big)t\nonumber\\ 
		&+& \nu\,\sum_{j \neq k} \left\|C_jB_k\right\| + \left\|T^1_{jk} \right\|\left\|I_1(t) \right\|_1.
		\end{eqnarray}
		Similarly,
		\begin{eqnarray}\label{J2}
		\left\|J_2(t)\right\|_1 &\leq& \gamma \sum_{j=1}^{2}g^{jj}\alpha^2_j\Big(\left\| B_j\right\|^2 + \left\|C_j\right\|^2\Big)t^3 \nonumber\\
		&+& \gamma\,\sum_{j\neq k}g^{jk} (\alpha^2_j\left\|B_k \right\|^2 + \alpha^2_k\left\| C_j\right\|^2)t^3\nonumber\\
		&+& \gamma\,\sum_{j \neq k}g^{jk}\left\| C_jB_k\right\|^2 + \left\|T^1_{jk} \right\|\left\|I_2(t) \right\|_1 ,
		\end{eqnarray}
	and 
		\begin{eqnarray}\label{J3}
		\left\|J_3(t)\right\|_1 &\leq&  \eta\,\sum_{j=1}^{2}g^{jj}\alpha^3_j\Big(\left\| B_j\right\|^3 + \left\|C_j\right\|^3\Big)t^4 \nonumber\\
		&+&\eta\,\sum_{j\neq k}g^{jk} (\alpha^3_j\left\|B_k \right\|^3 + \alpha^3_k\left\| C_j\right\|^3)t^4\nonumber\\
		&+& \eta\,\sum_{j \neq k}g^{jk}\left\| C_jB_k\right\|^3 + \left\|T^1_{jk} \right\|\left\|I_3(t) \right\|_1 ,
		\end{eqnarray}
		where $(\gamma, \eta)\in\mathbb{R}_{+}^{*}$.
	\end{itemize}
	Thus, 
	\begin{eqnarray}
	V(\mathcal{L}^m) - V(\mathcal{L}_0^m) &=& \lim_{t \rightarrow 0+ }tTr \Big( \left\|J_1(t)\right\|_1 + \left\|J_2(t)\right\|_1+    \left\|J_3(t)\right\|_1\Big)\nonumber\\
	&=& 0.
	\end{eqnarray}
	Therefore, for $r_j\in\mathcal{D}_{\theta}$, the volume is invariant under perturbation from $\mathcal{L}_0^m$ to $\mathcal{L}^m$.
	\item The integrated scalar curvature is given by (\ref{is})
	and according to 	
	the perturbation of the volume from $\mathcal{L}^m_0$ to $\mathcal{L}^m$, we have
	\begin{equation}
	s(\mathcal{L}^m) -s(\mathcal{L}^m_0) = \frac{1}{6} \lim_{t\longrightarrow 0+}[ Tr( e^{t\mathcal{L}^m}-e^{t\mathcal{L}^m_0})].
	\end{equation}
	\begin{itemize}
		\item Let us start with the case  $r_j\in\mathcal{D}_{\theta}^{\infty},$  where $j\in\{ 1,2\}$.
		
		Using the relation (\ref{j2}) and \cite{CGS}, we deduce that
		 $TrJ_2(t) \longrightarrow 0$ as $t\longrightarrow 0^+.$
		
		Since $J_3(t)$ vanishes, thus the computation of $s(\mathcal{L}^m) -s(\mathcal{L}^m_0)$ concerns only $J_1(t)$.
		
		According to the Proposition \ref{p1}, we have
		\begin{eqnarray}\label{s1}
		s(\mathcal{L}^m) -s(\mathcal{L}^m_0) 
		&=& \frac{1}{6}\lim_{t\longrightarrow 0+} t\Big(Tr\big(T^1_{jk}e^{t\mathcal{L}^m_0}\big) +  Tr\big(T^2_{jk}e^{t\mathcal{L}^m_0}\big)\Big).
		\end{eqnarray}
		Let us now compute $Tr\big(T^1_{jk}\,e^{t\mathcal{L}^m_0}\big)$ and $Tr\big(T^2_{jk}e^{t\mathcal{L}^m_0}\big)$, for all $t>0$.
		
		For $j=1$, $k=2$,  and $r\in\mathcal{D}^{\infty}_{\theta}$, we have $Tr\big(d_{r_1}d_1\,e^{t\mathcal{L}^m_0}\big)=Tr\big(d_{r_1}d_2\big)=0,$ 
		Then, for  $(j,k)\in\{1,2\}$, we deduce $Tr\big(T^1_{jk}e^{t\mathcal{L}^m_0}\big)=0.$
		
		Moreover,
		\begin{eqnarray}
		Tr\big(T^2_{jk}e^{t\mathcal{L}^m_0}\big) &=& Tr\Big(\sum_{j=1}^{2}g^{jj}(d^2_{r_j} - 2\pi i G_jd_{r_j})e^{t\mathcal{L}^m_0}\Big)\nonumber\\
		&-& Tr\Big(\sum_{j\neq k}g^{jk}(2\pi i G_kd_{r_j} 
		+ 2\pi i G_jd_{r_k})e^{t\mathcal{L}^m_0}\Big).
		\end{eqnarray} 
For $j\neq k$ and $r\in\mathcal{D}^{\infty}_{\theta}$, we have $Tr\Big(d_{r_j}e^{t\mathcal{L}^m_0}\Big)=Tr\Big(d_{r_k}e^{t\mathcal{L}^m_0}\Big) = 0.$
		Thus, $Tr\big(T^1_{jk}e^{t\mathcal{L}^m_0}\big) = Tr\Big(\displaystyle\sum_{j=1}^{2}g^{jj}d^2_{r_j}e^{t\mathcal{L}^m_0}\Big)$
		and according to Lemma \ref{l1}, the
		relation (\ref{s1}) becomes
		\begin{eqnarray}
		s(\mathcal{L}^m) -s(\mathcal{L}^m_0) &=& \frac{1}{6}\lim_{t\longrightarrow 0+} tTr\Big(\sum_{j=1}^{2}g^{jj}d^2_{r_j}e^{t\mathcal{L}_0}e^{t T^0_{jk}}\Big),
		\end{eqnarray}
		where $k\in\{1,2\}$.
		We perform the same computation as in \cite{CGS}, and deduce that  there exists $K\in\mathbb{R}^*_+$ such that
		\begin{equation}
		s(\mathcal{L}^m) -s(\mathcal{L}^m_0) \geq K.
		\end{equation}
		\item For $r_j\in\mathcal{D}_{\theta},$  with $j\in\{ 1,2\}$, we have
		\begin{eqnarray}
		s(\mathcal{L}^m) -s(\mathcal{L}^m_0) &=& \frac{1}{6}\lim_{t \longrightarrow 0^+} Tr\Big(J_1(t) + J_2(t) + J_3(t)\Big).
		\end{eqnarray}
		Furthermore, according to the relation (\ref{J1}), we get
		\begin{eqnarray}
		\lim_{t \longrightarrow 0^+}TrJ_1(t) 
		&=& \nu\,\sum_{j \neq k}g^{jk} \left\|C_jB_k\right\| + \left\|T^1_{jk} \right\|\left\|I_1(t)\right\|_1.
		\end{eqnarray} 
From \cite{CGS}, it follows that  
		$\displaystyle\lim_{t \longrightarrow 0^+}TrJ_1(t) = \nu\,\sum_{j \neq k}g^{jk} \left\|C_jB_k\right\|.$ 
		Similarly, we obtain
		\begin{equation}
		\lim_{t \longrightarrow 0^+}TrJ_2(t) = \gamma\,\sum_{j \neq k}g^{jk} \left\|C_jB_k\right\|^2
		\end{equation}
		and
		\begin{equation}
		\lim_{t \longrightarrow 0^+}TrJ_3(t) = \eta\,\sum_{j \neq k}g^{jk} \left\|C_jB_k\right\|^3.
		\end{equation}
		Thus, 
		\begin{eqnarray}
		s(\mathcal{L}^m) -s(\mathcal{L}^m_0) &=& \sum_{j \neq k}g^{jk}\Big(\nu\, \left\|C_jB_k\right\| + \gamma\, \left\|C_jB_k\right\|^2
		+ \eta\, \left\|C_jB_k\right\|^3 \Big),
		\end{eqnarray}
	\end{itemize}
	and, therefore, the result holds.
	$\cqfd$
\end{enumerate}
\section{Spectral triple on $\mathcal{D}^{\infty}_{\theta}$}
\subsection{Volume form on$\mathcal{D}^{\infty}_{\theta}$}
Now, we investigate  the invariance of the volume form under  the  perturbation from $\mathcal{L}^m_0$ to $\mathcal{L}^m.$  {We denote by $T_1$ the self adjoint operator with a compact resolvent $\tilde{T}_1 = T_1|\mathcal{N}(T_1)^{\perp},$ where $\mathcal{N}(T_1)$ is the kernel of $T_1$.}
\begin{definition}\cite{Con}
	The volume form on $\mathcal{D}_{\theta}^{\infty}$ is a linear functional given by:
	\begin{equation}\label{e41}
	v(x):= \frac{1}{2}{Tr}_w (x|\tilde {D}|^{-2} P),
	\end{equation}
	where, ${Tr}_w $ is the Dixmier trace, $D$ an operator, $x\in\mathcal{D}^{\infty}_{\theta}$,  and $P$ is the projection on $\mathcal{N}(T)^{\perp}$.
\end{definition}
\begin{lemma}\cite{S}\label{vf}
	Let $T_1$ be a self-adjoint operator with a compact resolvent such that ${\tilde {T}_1}^{-1} $
	is Dixmier trace-able. Then, for $x \in \mathcal{D}^{\infty}_{\theta} $ and every $b \in \rho (T_1),$ \begin{equation}\label{e43}
	{Tr}_w ( x {\tilde {T}_1}^{-1} P) = {Tr}_w ( x {(T_1-z)}^{-1}),
	\end{equation}
	{where $\rho$ is a resolvent}.
\end{lemma}
\begin{lemma}
	The operator $\partial: \mathcal{D}^{\infty}_{\theta}\longrightarrow \mathcal{D}^{\infty}_{\theta}$  defined as follows: 
	\begin{equation}
	\partial=\sum_{j=1}^{2}(d_j -2\pi i G_j),
	\end{equation}
	is a self-adjoint operator on $\mathcal{D}^{\infty}_{\theta}$.
\end{lemma}
{\it Proof:}
Since $d:\mathcal{D}^{\infty}_{\theta} \longrightarrow \mathcal{D}^{\infty}_{\theta}$ is self-adjoint, thus, we deduce that for $j\in\{ 1,2 \}$, $d_j- 2\pi iG_j : \mathcal{D}^{\infty}_{\theta} \longrightarrow \mathcal{D}^{\infty}_{\theta}$ is self-adjoint.
Therefore, we get the result. $\cqfd$
\begin{definition}
	\begin{enumerate}
		\item {The unperturbed  magnetic operator $D^m_0 :\mathcal{D}^{\infty}_{\theta} \longrightarrow \mathcal{D}^{\infty}_{\theta}$ is given by:
		\begin{equation}\label{do}
		D^m_0 = \left( \begin{array} {cc}
		0 &  d^m_1 + i\,d^m_2   \\
		 d^m_1 - i\,d^m_2 & 0
		\end {array} \right).
		\end{equation}}
		\item For $r \in \mathcal{D}_{\theta}^\infty$, the perturbed magnetic  operator $D^m :\mathcal{D}^{\infty}_{\theta} \longrightarrow \mathcal{D}^{\infty}_{\theta}$ is defined as follows:
		\begin{equation}\label{dm}
		D^m = D^m_0 + \left( \begin{array} {cc}  0 & d_r  \\
		d_{r^*}   & 0
		\end {array} \right). 
		\end{equation}
	\end{enumerate}
\end{definition}
\begin{definition}
	The unperturbed and perturbed volume forms {associated with a magnetic Laplacian} are given by: 
	\begin{equation}\label{e42}
	v^m_0 (x) := \frac{1}{2}{Tr}_w(x| {\tilde{D}^m}_0|^{-2}P)\quad\mbox{and}\quad v^m(x):= \frac{1}{2}{Tr}_w(x{|\tilde{D}^m|}^{-2}P),
	\end{equation}
	where  $x \in \mathcal{D}^{\infty}_{\theta}$.	
\end{definition}
{\begin{remark}
		\begin{enumerate}
			\item The unperturbed magnetic operator is given by:\begin{equation}\label{r1}
			D^m_0 = D_0 + G
			\end{equation}
				where
				\begin{equation}\label{go}
				G = -2\pi\,i\left( \begin{array} {cc}
				0 &  G_1 + i\,G_2   \\
				G_1 - i\,G_2 & 0
				\end {array} \right).
				\end{equation}
				\item The perturbed magnetic operator is also written as follows: \begin{equation}
				D^m = D + G. 
				\end{equation}
				\item When $G_j =0,$ for $j\in\{1,2\},$ we obtain the unperturbed and perturbed operators defined by Chakraborty and co-workers\cite{CGS}.			
		\end{enumerate}
	\end{remark}}
\begin{theorem}
	The volume form {associated with the magnetic Laplacian} is invariant under the perturbation from $\mathcal{L}^m_0$ to $\mathcal{L}^m.$
\end{theorem}
\textit{Proof:}
{It is obvious to have:
\begin{equation}\label{sp1}
(D^m_0)^2 =  \left( \begin{array} {cc}  \tilde{\mathcal{L}}^m_0 & 0  \\
0   & \hat{\mathcal{L}}^m_0
\end {array} \right),
\end{equation}
where $\tilde{\mathcal{L}}^m_0$ and $\hat{\mathcal{L}}^m_0$ are given, respectively, by:
\begin{equation}
\tilde{\mathcal{L}}^m_0 = \mathcal{L}^m_0 + i\,d^m_2\,d^m_1 - i\,d^m_1\,d^m_2 - \sum_{j\neq k}g^{jk}d^m_j\,d^m_k
\end{equation}
and
\begin{equation}
\hat{\mathcal{L}}^m_0 = \mathcal{L}^m_0 + i\,d^m_1\,d^m_2 - i\,d^m_2\,d^m_1  - \sum_{j\neq k}g^{jk}d^m_j\,d^m_k.
\end{equation}
Furthermore,
\begin{equation}
(D^m)^{2}=  \left( \begin{array} {cc}  \mathcal{L}^m_1 & 0  \\
0  & \mathcal{L}^m_2
\end {array} \right),
\end{equation}
with
\begin{equation}
\mathcal{L}^m_1 = \tilde{\mathcal{L}}^m_0 + (d^m_1 + i\,d^m_2)d_{r^*} 
+ d_r\,(d^m_1 - i\,d^m_2) + d_rd_{r^*}. 
\end{equation}
and
\begin{eqnarray}
\mathcal{L}^m_{2}= \hat{\mathcal{L}}^m_0 + d_{r^*}\,(d^m_1 + i\,d^m_2) 
+ (d^m_1 - i\,d^m_2)\,d_r + d_{r^*}d_r.
\end{eqnarray}
Let us consider $P_1$ and $P_2$, two projections on $\mathcal{N}(\mathcal{L}^m_1)^{\perp}$ and $\mathcal{N}(\mathcal{L}^m_2)^{\perp},$ respectively. By Theorem $5.3$ (see \cite{S}), $\mathcal{L}^m_{1}$ and $\mathcal{L}^m_{2}$ have compact resolvent and using the above Lemma \ref{vf} for $Imy\neq 0$, we have
\begin{equation}
v^m_{0}(x) = {Tr}_w (x{(-\tilde{\mathcal{L}}^m_0 -y)}^{-1}	+ x{(-\hat{\mathcal{L}}^m_0 -y)}^{-1}).
\end{equation}
Since $
{(-\mathcal{L}^m_i-y)}^{-1}-{(-\mathcal{L}^m_0 -y)}^{-1}$ is a trace class for $ i=1,2$, then
\begin{eqnarray}
v^m(x) &=& {Tr}_w \big(x {(-\mathcal{L}^m_1 -y)}^{-1}	+ x {( -\mathcal{L}^m_2 -y )}^{-1}\big)\nonumber\\
&=& {Tr}_w \big(x {(-\mathcal{L}^m_1 -y)}^{-1}	+ x{( -\tilde{\mathcal{L}}^m_0 -y )}^{-1}\big)\nonumber\\
&+& {Tr}_w \big(x {(-\mathcal{L}^m_2 -y)}^{-1}	+ x{( -\hat{\mathcal{L}}^m_0 -y )}^{-1}\big)\nonumber\\
&+&{Tr}_w \big(x {(-\tilde{\mathcal{L}}^m_0 -y)}^{-1}	+ x{( -\hat{\mathcal{L}}^m_0 -y )}^{-1}\big)\nonumber\\
&=& v^m_0(x).
\end{eqnarray}
Therefore, the volume form  associated with the magnetic Laplacian is invariant under our perturbation.  $\cqfd$}
\subsection{Spectral triple}
Let us recall some basic notions concerning the spectral triples.
\begin{definition}\cite{S}
	An even spectral triple for a $*-$ algebra $\mathcal{A}$ is a triple $\big( \pi, \mathcal{H}, \mathcal{D}\big)$ together with a $\mathbb{Z}_2$ grading $\gamma$ on $\mathcal{H}$ such that
	\begin{enumerate}
		\item The map $\pi : \mathcal{A} \to \mathcal{L}(\mathcal{H})$ is a $*-$ representation such that $\pi(x)\gamma = \gamma\pi(x)$ for $x\in\mathcal{A}$.
		\item The operator $D$ is an unbounded operator with compact resolvent such that $D\gamma = -\gamma D$.
		\item  The commutator $[D, \pi(x)]$ is bounded for every $x\in \mathcal{A}$.
	\end{enumerate}
\end{definition}
\begin{definition}
	Let $\mathcal{B}$ be a $*-$ algebra, $\mathcal{K}$ a Hilbert space, and $D$ an operator. Then,  the spectral triples $\big(\mathcal{B}_1, \mathcal{K}_1, D_1 \big)$ and $\big(\mathcal{B}_2, \mathcal{K}_2, D_2 \big)$ are unitarily equivalent if there exists a unitary operator $V$ and a representation $\pi_l$, for $l\in\{1,2\}$ defined as follows:
	$$V: \mathcal{K}_1 \longrightarrow \mathcal{K}_2 \quad \mbox{such that}\quad D_2 = VD_1V^*$$ and
	$$ \pi_l :\mathcal{B}_2\longrightarrow \mathcal{K}_l\quad \mbox{,}\quad \pi_2(x)= V\pi_1(x)V^*,$$
 respectively,
\end{definition}
In the rest of this work, we call the magnetic even spectral triple, the even spectral triple associated with the magnetic Laplacian.

The next Lemma characterizes the magnetic even spectral triple. We consider the Hilbert space $\mathcal{H}=L^2(\varphi) \oplus L^2(\varphi).$
\begin{lemma}
	Let $\Gamma$ be the grading operator. Then,
	\begin{enumerate}
		\item The triple $\big( \mathcal{D}_{\theta}^{\infty}, L^2(\varphi) \oplus L^2(\varphi), D^m_0\big)$ is an unperturbed magnetic even spectral triple.
		\item The triple $\big( \mathcal{D}^{\infty}_{\theta}, L^2(\varphi) \oplus L^2(\varphi), D^m_0\big)$ is a perturbed magnetic even spectral triple.
	\end{enumerate}
\end{lemma}
{\it Proof:} For the proof, we consider the grading operator given as follows: 
$
\Gamma = \left(
\begin{array}{cc}
I & 0 \\
0 & -I\\
\end{array}
\right)$
and the $*-$ representation $\pi(x)= x$ where $x\in\mathcal{D}^{\infty}_{\theta}$.
\begin{enumerate}
	\item It is obvious to see that $x\Gamma = \Gamma x$, $\Gamma^* = \Gamma = \Gamma^{-1}$ and $\Gamma\,G=-\Gamma\,G.$
	Since $D^m_0 = D_0 + G$, and according to \cite{S}, we have, $	\Gamma\,D^m_0=-D^m_0\,\Gamma.$ 
	Furthermore, the dimension of $kerD^m_0$ is $2,$
	and according to the relation (\ref{sp1}), we conclude that the unperturbed magnetic $D^m_0$ has a compact resolvent.
	
	In addition, 
	for $x\in\mathcal{D}^{\infty}_{\theta},$ $[G,x]$ is bounded, and using \cite{CGS}, we conclude that $	[D^m_0, x]=  [D_0, x] + [G, x]$ is bounded
	\item Similarly, we obtain the result.
\end{enumerate}
$\cqfd$

From \cite{Con}, the  next definition is formulated.
\begin{definition}
	Let $\Omega^1\big(\mathcal{D}_{\theta}^{\infty}\big)$ be the universal space of $1-$ forms. Then, the representation $\pi : \Omega^1\big(\mathcal{D}_{\theta}^{\infty}\big) \longrightarrow \mathcal{K}$ is defined as follows:
	\begin{equation}\label{pi}
	\pi(x) = x \quad\mbox{and}\quad \pi\big(\delta^m(x)\big) = [D^m, x].
	\end{equation}
\end{definition}
{\begin{lemma}
	Let  $D^m_0$ and $D^m$ two operators  both mapping to $\mathcal{D}^{\infty}_{\theta}$ from $\mathcal{D}^{\infty}_{\theta}$. Then, the following relations hold: 
	\begin{enumerate}
		\item 
		\begin{equation}\label{dom}
		D^m_0= i\big( d_1\gamma_1 + d_2\gamma_2\big) - 2\pi\,i\big( G_1\gamma_1 + iG_2\gamma_2\big).
		\end{equation}
		\item
		\begin{equation}\label{pix}
		[D^m,\,x] = [D,\,x] - 2\pi\,i\big( G_1\gamma_1 + iG_2\gamma_2\big),
		\end{equation}
	\end{enumerate}
	where $\gamma_1$, $\gamma_2$ are the $2\times2$ Clifford matrices.
\end{lemma}
{\it Proof:}
\begin{enumerate}
	\item According to the relation (\ref{r1})
	and using \cite{CGS}, we have $D^m_0 = i\big(\gamma_1d_1 + \gamma_2d_2\big) - G.$ Moreover, the matrix $G$ is expressed as $
	G= 2\pi\,iG_1\gamma_1 - 2\pi G_2\gamma_2.$
	 Thus, the relation (\ref{dom}) holds.
	\item 
	Using the relations (\ref{dm}), (\ref{r1}), and after computation, we get the result.
\end{enumerate}
$\cqfd$ }

Now, we study the unitarity of the unperturbed and perturbed magnetic spectral triples.
\begin{definition}\cite{HGPG}
	Let $\big( \pi_1, \mathcal{H}_1, \mathcal{D}_1\big)$ and  $\big( \pi_2, \mathcal{H}_2, \mathcal{D}_2\big)$ be two even spectral triples such that $\pi = \pi_1 + \pi_2$, $\mathcal{H}= \mathcal{H}_1 \oplus \mathcal{H}_2$ and $D= D_1 + D_2$. Then, $\big( \pi, \mathcal{H}, \mathcal{D}\big)$ is an even spectral triple, the so called the direct sum of $\big( \pi_1, \mathcal{H}_1, \mathcal{D}_1\big)$ and  $\big( \pi_2, \mathcal{H}_2, \mathcal{D}_2\big)$.	
\end{definition}

We arrive at the following Proposition.
\begin{proposition}
	Let $X\in\mathcal{D}_{\theta}.$  Then, for $r = X^{n_1},$ and $n_1\in\mathbb{Z},$ we have:
	\begin{enumerate}
		\item $\pi(\Omega^1) = \mathcal{D}^{\infty}_{\theta}\oplus\mathcal{D}^{\infty}_{\theta}$.
		\item $\Omega^2(\mathcal{D}^{\infty}_{\theta})=0.$
	\end{enumerate}
\end{proposition}
{\it Proof:}
{\begin{enumerate}
	\item According to the relations (\ref{pi}) and (\ref{pix}), we get 
	\begin{equation}
	\pi\big(\delta^m(x)\big) = \pi\big(\delta(x)\big) - 2\pi\,i\big( G_1\gamma_1 + iG_2\gamma_2\big).
	\end{equation}
	Setting $\Omega^1(\mathcal{D}^{\infty}_{\theta})=- 2\pi\,i\big( G_1\gamma_1 + iG_2\gamma_2\big)$
	and using the Lemma \ref{l1}, we obtain 
	\begin{eqnarray}
	\Omega^1_{D^m}(\mathcal{D}^{\infty}_{\theta})
	&=& \Omega^1_{D}(\mathcal{D}^{\infty}_{\theta}) + \Omega^1_{T_{jk}}(\mathcal{D}^{\infty}_{\theta}).
	\end{eqnarray}
	Since $\Omega^1_{D}(\mathcal{D}^{\infty}_{\theta})=\mathcal{D}^{\infty}_{\theta} \oplus \mathcal{D}^{\infty}_{\theta},$ performing the same calculation as in \cite{CGS}, we deduce that $\Omega^1_{T_{jk}}(\mathcal{D}^{\infty}_{\theta})= \mathcal{D}^{\infty}_{\theta} \oplus \mathcal{D}^{\infty}_{\theta}.$ 
	Therefore, $	\Omega^1_{D^m}(\mathcal{D}^{\infty}_{\theta})=\mathcal{D}^{\infty}_{\theta} \oplus \mathcal{D}^{\infty}_{\theta}.$ 
	\item It is obvious.
	$\cqfd$
\end{enumerate}}
Thus, we obtain the following result. 
\begin{proposition}
	For $r = X^{n_1}$ with $n_1\in\mathbb{Z}$ and $X\in\mathcal{D}_{\theta}$, the unperturbed magnetic and perturbed magnetic spectral triples $\big(\mathcal{D}^{\infty}_{\theta},  L^2(\varphi) \oplus L^2(\varphi), D^m_0\big)$ 
	and  $\big(\mathcal{D}^{\infty}_{\theta},  L^2(\varphi) \oplus L^2(\varphi), D^m\big)$ are not unitarily equivalent.
\end{proposition}
{\it Proof:}
Using Lemma \ref{l1} and Theorem $(4.4)$\cite{CGS}, it suffices to study the unitarity of the even spectral triples $\big(\mathcal{D}^{\infty}_{\theta}, L^2(\varphi) \oplus L^2(\varphi), T^0_{jk}\big)$ and  $\big(\mathcal{D}^{\infty}_{\theta}, L^2(\varphi) \oplus L^2(\varphi), T_{jk}\big).$ According to the above Proposition, $\Omega^2(D^m)=0$, but 
$\Omega^2(D^m)=\Omega^2(D) + \Omega^2(T_{jk})$. Thus, using \cite{S}, we deduce that $\Omega^2_{T_{jk}}(\mathcal{D}^{\infty}_{\theta})=0$. Since $\Omega^2_{T^0_{jk}}(\mathcal{D}^{\infty}_{\theta})= \mathcal{D}^{\infty}_{\theta}$, thus, the result holds.
$\cqfd$

Let $\mathcal{K}^m$ be the vector space of all magnetic derivations $d^m: \mathcal{D}_{\theta}^{\infty} \longrightarrow \mathcal{D}_{\theta}^{\infty}$. $\mathcal{K}^m$ is identified as the space of derivations given by \cite{BEJ},  i.e the space of the form\\ {$\left\lbrace \displaystyle \sum_{j=1}^{2}c_jd^m_j + [r, .], r\in\mathcal{D}_{\theta}^{\infty},~c_j \in \mathbb{C}.\right\rbrace .$} 

Let $\mathcal{B}$ be any normed $\mathcal{D}_{\theta}^{\infty} -$ module. For $\delta^m \in\mathcal{K}^m$, we consider the contraction $C_{\delta^m} : \mathcal{B}\otimes \mathcal{K}^m \longrightarrow \mathcal{B}$.
\begin{definition}
	The magnetic connection is a complex linear map defined by:
	$\nabla^m: \mathcal{B}\otimes \mathcal{K}^m \longrightarrow \mathcal{B}$ such that
	\begin{equation}
	C_{\delta^m}\nabla^m(\xi x)= C_{\delta^m}\nabla^m(\xi) x + \xi\, \delta^m(x),
\quad \forall\, \delta^m\in\mathcal{K}^m. 
	\end{equation}
\end{definition}
In addition, we define a magnetic connection as follows:
\begin{equation}
C_{\delta^m}\nabla^m(\xi x) = C_{\delta}\nabla(\xi x) - \xi\,\alpha_j(x),
\end{equation}
where $\alpha_j = 
2\,\pi\,i\displaystyle\sum_{j=1}^{2}c_j\,G_j.$

We assume that the maps $\nabla^m_j: \mathcal{B} \longrightarrow \mathcal{B} $ for $j\in\{1,2\}$ are given by the following relation
\begin{equation}
\nabla^m_j(\xi\,x) = \nabla^m_j(\xi)\,x + \xi\,d^m_j(x).
\end{equation}
\begin{proposition}
	The map $\nabla^m : \mathcal{B} \longrightarrow \mathcal{B} $ given by: 
	\begin{equation}
	C_{\delta^m}\nabla^m(\xi) = \displaystyle\sum_{j=1}^{2}\nabla^m_j \otimes d^m_j - \sum \xi\,X^{n_1}\,Y^{n_2}\otimes \delta_{n_1n_2}
	\end{equation}
	is a magnetic connection if  
	 $ \nabla_j  \otimes\,G_j= \xi\big(G_j\otimes\,d_j - G_j\otimes\,G_j\big)$. 
\end{proposition}
{\it Proof:} Since $d_j^m = d_j - \beta_j$, where $\beta_j = 2\pi\,i\,G_j$,  thus
\begin{eqnarray}\label{c1}
\displaystyle\sum_{j=1}^{2}\nabla^m_j \otimes d^m_j 
&=& \displaystyle\sum_{j=1}^{2}\nabla^m_j \otimes d_j - 2\pi\,i\,\displaystyle\sum_{j=1}^{2}\nabla^m_j \otimes G_j.
\end{eqnarray}
Moreover, 
\begin{equation}\label{a1}
\nabla^m_j(\xi\,x) = \nabla_j(\xi\,x) - \xi\,\beta_j(x).
\end{equation}
Thus, using the relation (\ref{c1}), we obtain

\begin{eqnarray}
\nabla^m(\xi) &=& \nabla(\xi) + \gamma(\xi),
\end{eqnarray}
where \begin{equation}
\gamma(\xi) = 2\pi\, i\,\sum_{j=1}^{2}\Big(\xi\,G_j\otimes\,d_j - \nabla_j \otimes\,G_j - \xi\,G_j\otimes\,G_j\Big).
\end{equation}
According to \cite{CGS}, the map $ \nabla $ is a connection.
  It follows that $\nabla^m$ is a connection. $\cqfd$

Considering the definition of the connection, we consider the following relations.
\begin{equation}\label{as}
\left \{
\begin{array}{l}
C_{\beta}\nabla=C_{d}\beta=0\\
\\
C_{d_r}\nabla=\nabla_r\\
\\
C_{d_j}\nabla=\nabla_j
\end{array}
\right .
\end{equation}
{We consider $\mathcal{L}(\mathcal{B}),$  the space of linear forms on $\mathcal{B}.$}
\begin{definition}
	The curvature $2-$ form $R: \mathcal{K}^m\,\otimes\,\mathcal{K}^m \longrightarrow \mathcal{L}(\mathcal{B})$ associated with the magnetic connection $\nabla^m$ is defined by:
	\begin{equation}\label{r2}
	R\big( \delta^m_1, \delta^m_2\big) = C_{[\delta^m_1, \delta^m_2]}\nabla^m - [C_{\delta^m_1}\nabla^m, C_{\delta^m_2}\nabla^m].
	\end{equation}
\end{definition}
\begin{theorem}
	The curvature $2-$ form associated with the magnetic connection is not invariant under the perturbation by inner derivation.
\end{theorem}
{\it Proof:} It  suffices to show the relation between $R\big ( d^m_1 , d^m_2\big)$ and $R\big ( \delta^m_1 , \delta^m_2\big)$.
Using the relations (\ref{as}) and (\ref{a1}), we obtain, respectively,  
\begin{eqnarray}
[C_{\delta_1}\nabla^m, C_{\delta_2}\nabla^m] 
&=& [C_{\delta_1}\nabla,\,C_{\delta_2}\nabla]
\end{eqnarray}
and 
\begin{eqnarray}
[C_{d^m_1}\nabla^m,\, C_{d^m_2 }\nabla^m]
&=&  [C_{d_1}\nabla,\, C_{d_2}\nabla] .
\end{eqnarray}
Moreover,
\begin{eqnarray}
[\delta^m_1,\,\delta^m_2] 
&=& [\delta_1,\,\delta_2]+ 2\pi\,i\big(G_2\delta_1-G_1\delta_2 + \psi_{21}\big),
\end{eqnarray}
\begin{eqnarray}
[d^m_1,\,d^m_2] 
&=& [d_1,\,d_2]+ 2\pi\,i\big(G_2\,d_1-G_1\,d_2 + \psi_{21}\big).
\end{eqnarray}
Thus, using once (\ref{a1}), we obtain
\begin{eqnarray}
C_{[\delta^m_1,\,\delta^m_2]}\nabla^m 
&=& C_{ [\delta_1,\,\delta_2]}\nabla + 2\pi\,i\big(G_2(\nabla_1+\nabla_{r_1}) - G_1(\nabla_2+\nabla_{r_2})\big),
\end{eqnarray}
and
\begin{eqnarray}
C_{[d^m_1,\,d^m_2]}\nabla^m 
&=& C_{[d_1,\,d_2]}\nabla + 2\pi\,i\big(G_2\nabla_1 - G_1\nabla_2\big).
\end{eqnarray}
Therefore, 
according to the invariance condition of the curvature $2-$form \cite{CGS,S}, we have
\begin{eqnarray}\label{ra}
R\big( \delta^m_1,\,\delta^m_2\big) &=& R \big( d_1,\,d_2\big)+ 2\pi\,i\big(G_2(\nabla_1+\nabla_{r_1}) - G_1(\nabla_2+\nabla_{r_2})\big) .
\end{eqnarray}
Futhermore,
\begin{eqnarray}\label{rb}
R\big( d^m_1,\,d^m_2\big) &=& R\big( d_1,\,d_2\big) + 2\pi\,i\big(G_2\nabla_1 - G_1\nabla_2\big).
\end{eqnarray}
Therefore, using the relations (\ref{ra}) and (\ref{rb}), we get
\begin{equation}
R\big(\delta^m_1 , \delta^m_2\big) = R\big( d^m_1,\,d^m_2\big) + 2\pi\,i\big(G_2\nabla_{r_1} - G_1\nabla_{r_2})\big).
\end{equation}
$\cqfd$
\section{Noncommutative $2$d-dimensional space}
In this section, we study the  magnetic quantum stochastic differential equation (mqsde) in $2d$-dimensional noncommutative  space and some properties of  its solution.
For more information on the geometry of the simplest of noncompact manifolds (Euclidean 2d-dimensional space), see \cite{CGS} and \cite{FG}.
\subsection{Basic tools}
For the integer $d\geq 1$, we consider $\mathcal{C}_0\big(\mathbb{R}^{2d}\big)$ as the $\mathcal{C}^*-$ algebra of complex valued continuous functions on $\mathbb{R}^{2d}$ wich vanish at infinity. The magnetic partial derivative on $\mathcal{C}_0\big(\mathbb{R}^{2d}\big)$ in the k-th direction is the operator $\partial^m_k = \partial_k - 2\pi\,i\,G_k$ with $k\in\{1,2,\cdots,2d \}$. 

The space of smooth complex  valued functions on $\mathbb{R}^{2d}$ with compact support is denoted by $\mathcal{B}^{\infty}_{c}:= \mathcal{C}^{\infty}_c\big(\mathbb{R}^{2d}\big)$. Let us consider the Hilbert space of square  integrable  functions, $L^2\big(\mathbb{R}^{2d}\big).$ Then, the symmetric linear magnetic operator which maps on $L^2\big(\mathbb{R}^{2d}\big)$ with the domain $\mathcal{B}^{\infty}_c$ is defined by $i\partial^m_k,$ $i\partial^m_k = i\partial_k + 2\pi\,G_k$.

We consider $\left\langle , \right\rangle $ as the Euclidean inner product of $\mathbb{R}^{d}$.
\begin{definition}\cite{CGS}
	The Fourier transform on the Hilbert space $L^2\big(\mathbb{R}^{2d}\big)$ is defined by:
	\begin{equation}\label{ft}
	\hat{g}(\epsilon) = (2\,\pi)^{-d}\int e^{i\,\left\langle \epsilon, u\right\rangle }g(u)\,du.
	\end{equation}
\end{definition}
The operator of multiplication by the function $\psi$ is defined by $M_{\psi},$ and setting $\widehat{M}_{\psi} = \mathcal{F}^{-1}\,M_{\psi}\,\mathcal{F}$, we get $i\partial^m_j= i\partial_j + \left\langle   \widehat{M}_{\psi}, T^0_{jk} \right\rangle.$
\begin{definition}
	The $2-$d dimensional magnetic operator is given by:
	\begin{equation}
	i\,\partial^m_j = i\,\partial_j + \widehat{M}_{f(x_{jk})}\quad\mbox{,}\quad (j,k)\in\{1,2
	\}
	\end{equation}
	where
	\begin{equation}
	f(x_{jk}) = \sum_{j=1}^{2}\left(\pi i G_j  x_j + 2\pi^2 G^2_j \right) + \sum_{j\neq k}(x_j - 2\pi\,i\,G_j)(x_k - 2\pi\,i\,G_k).
	\end{equation}
\end{definition} 
We assume that the differential operator $\displaystyle\sum_{j,k=1}^{2d}(d_j - 2\pi\,i\,G_j)(d_k - 2\pi\,i\,G_k)$ is the restriction of the magnetic Laplacian $\Delta^m$ on $\mathcal{B}^{\infty}_c$.
\begin{definition}
	The continuous groups of unitaries in  $\tilde{\mathcal{H}}:= L^2\big(\mathbb{R}^d\big),$  the space of square  integrable functions on $\mathbb{R}^d,$ are defined as follows:
	\begin{equation}
	\left \{
	\begin{array}{l}
	\displaystyle
	(X_{\alpha}g)(u) = g(u + \alpha) \\
	(Y_{\mu}g)(u)= \exp\,i\,\left\langle u, \mu \right\rangle\,g(u),
	\end{array}
	\right .
	\end{equation}
	where $\alpha$ , $\mu$, $u \in\mathbb{R}^d$, and $g\in\mathcal{C}^{\infty}_c\big(\mathbb{R}^{d}\big)$.
\end{definition} 
It is  straightforward to see that $X_{\alpha}$ and $Y_{\mu}$ verify the following relations:
\begin{equation}\label{fr}
\left \{
\begin{array}{l}
\displaystyle
X_{\alpha_1}\,X_{\alpha_2} = X_{\alpha_1 + \alpha_2}\\
Y_{\mu_1}\,Y_{\mu_2}= Y_{\mu_1 + \mu_2}\\
X_{\alpha_1}\,Y_{\mu_1} =\exp\,i\,\left\langle \alpha_1, \mu_1 \right\rangle\ Y_{\mu_1}\,X_{\alpha_1}
\end{array}
\right .
\end{equation}
and the equation (\ref{fr}) takes the form: 
\begin{equation}
Z_u\,Z_v = Z_{u+v}\,e^{(i/2)f(u,v)},
\end{equation} 
with $f(u,v) = u_1.v_2 - u_2.v_1$ such that $u=(u_1,u_2)$ and $v=(v_1,v_2)$.

Let $\mathcal{B}\big(\tilde{\mathcal{H}}\big)$ be the space of bounded operators in $\tilde{\mathcal{H}}$. For $\hat{g}\in L^1\big(\mathbb{R}^{2d}\big)$, we set
\begin{equation}
a(g)= \int_{\mathbb{R}^{2d}}\hat{g}(u)Z_udu.
\end{equation}
We denote by $\mathcal{B}^{\infty}$  the $*$- algebra of elements of the form $\left\lbrace a(g)\quad \mbox{s.t}\quad g\in\mathcal{C}^{\infty}_c\big(\mathbb{R}^d\big)\right\rbrace,$ and $\mathcal{B}$ the $\mathcal{C}^{*}$-algebra generated by $\mathcal{B}^{\infty}.$ For $(g, h)\in\mathcal{C}^{\infty}_c\big(\mathbb{R}^d\big)$, we set
\begin{equation}
(\widehat{g\odot h})(u) = \int \hat{g}(u-v)\hat{h}(v)e^{(i/2)f(u,v)}dv\quad\mbox{,}\quad g^{\mathbb{N}}(u)=\bar{g}(-u).
\end{equation}
Using the relation (\ref{fr}), it is obvious to verify that 
\begin{equation}
a(g)a(h) = a(g\odot h)\quad\mbox{and}\quad a(g)^*= a(g^{\mathbb{N}}).
\end{equation}
\begin{corollary}\cite{FG}
	The linear functional $\phi$ on $\mathcal{B}^{\infty}$ defined by
	\begin{equation}
	\phi(a(g)) = (2\pi)^{-d}\int g(u)du
	\end{equation}
	is a faithful trace on $\mathcal{B}^{\infty}$.
\end{corollary}
\begin{definition}
	The magnetic $2$d-parameter group of automorphisms of $\mathcal{B}$ is given by
	\begin{equation}
	\psi_{\alpha}(a(g)) = \int_{\mathbb{R}^{2d}}e^{i\left\langle \alpha, u\right\rangle }\hat{g}(u)Z_udu,
	\end{equation}
	where $\alpha\in\mathbb{R}^{2d}$ and $g\in\mathcal{C}^{\infty}_c\big(\mathbb{R}^{2d}\big).$
\end{definition}
It is straightforward to see that for $a(g)\in\mathcal{B}^{\infty}$, the map $\alpha \longrightarrow \psi_{\alpha}(a(g))$ is smooth.
\begin{definition}
	For $j\in\{1,\cdots,2d\}$, the magnetic derivation is defined as follows:
	\begin{equation}
	\delta^m_j(a(g)) = a(\partial_j(g)) - 2\pi\,i\,a(G_j(g)),
	\end{equation}
	with  $g\in \mathcal{C}^{\infty}_c\big(\mathbb{R}^{2d}\big).$
\end{definition}

Let us now look for  the case of a Riemannian manifold. We study the volume form associated with the magnetic $2$d-dimensional Laplacian $\Delta^m$. Let us consider  $\mathcal{T}^m_t$ as the contractive $C_0$ semigroup generated by the magnetic Laplacian, 
and  called the magnetic heat semigroup on $\mathbb{R}^{2d}$. It is easy to verify that, for $(j,k)\in\{1,\cdots,2d\}$,  $\mathcal{T}_t^m = \mathcal{T}_t\,e^{(t/2)T^0_{jk}},$ where $\mathcal{T}_t$ is the classical heat semigroup.
\begin{proposition}\label{p2}
	Let $\mathcal{T}^m_t$ be the magnetic heat semigroup and $M_g$ be the multiplication operator by the function $g$. Then, 
	\begin{enumerate}
		\item $M_g\,\mathcal{T}^m_t$ is a trace-class.
		\item $Tr\big(M_g\,\mathcal{T}^m_t\big) = (2\pi)^d\,t^{-d} f(G_1,G_2).$
	\end{enumerate}
\end{proposition}
{\it Proof:}
\begin{enumerate}
	\item 
	Since $M_g\mathcal{T}^m_t = \mathcal{F}\,M_g\,\mathcal{F}^{-1}\,M_{e^{-(t/2)\sum x^2_j + f(x_{jk})}},$ then,
	\begin{equation}
	Tr\big(M_g\mathcal{T}^m_t\big) = Tr\Big(\mathcal{F}\,M_g\,\mathcal{F}^{-1}\,M_{e^{-(t/2)\sum x^2_j + f(x_{jk})}}\Big),
	\end{equation}
	and the kernel of the integral operator is given by 
	\begin{equation}
	K^m_t(u,v) = \hat{g}(u-v)e^{-(t/2)w(v)},
	\end{equation}
	with $w(v)= \sum v^2_j + f(v_{jk}).$ It is easy to see that $K^m_t(u,v)$ is continuous in $u$ and $v$ and  $\int \left|K^m_t(u,u) \right|\,du < \infty .$ Following \cite{GK}, we deduce the result.\\
	\item \begin{eqnarray}
	Tr\big(M_g\mathcal{T}^m_t \big) &=& \int K^m_t(u,u)\,du\nonumber\\
	&=& \hat{g}(0)e^{-2t\pi^2G_1G_2}\int e^{-(t/2)w(u)}du.
	\end{eqnarray} 
	Moreover,  $w(u)= \big(u_1 + \tau\big)^2 +\big(u_2 + \sigma\big)^2 - \sigma^2  - \tau^2$ where 
	 $\tau = \frac{1}{2}\pi\,iG_1 - \pi\,iG_2 $ and 
 $\sigma = \frac{1}{2}\pi\,iG_2 - \pi\,iG_1$. Setting $f(G_1,G_2)= \exp\big((t/2)(4\pi^2G_1G_2 + \sigma^2 + \tau^2)\big),$
	we obtain
	\begin{eqnarray}
	Tr\big(M_g\mathcal{T}^m_t \big) &=& \hat{g}(0)f(G_1,G_2)\int e^{-(t/2)\big((u_2 + \sigma)^2 + (u_1 + \tau)^2\big)}du_1du_2\nonumber\\
	&=& f(G_1,G_2)Tr\big(M_g\mathcal{T}_t\big).
	\end{eqnarray}
	According to \cite{S}, we get
	\begin{equation}
	Tr\big(M_g\mathcal{T}^m_t \big) =(2\pi)^d\,t^{-d} f(G_1,G_2).
	\end{equation}
\end{enumerate}
$\cqfd$

In particular, we define the volume as follows:
\begin{equation}
v^m(g):= t^dTr\big(M_g\mathcal{T}^m_t \big).
\end{equation}
As in previous section, we give the analogue expression of the volume form using the Dixmier trace.
\begin{lemma}\cite{CGS}
	For $h,g_1 \in\,L^p\big(\mathbb{R}^2\big)$ with $2 \leq p < \infty $, then $M_h\widehat{M}_{g_1}$ is a compact operator in $L^p\big(\mathbb{R}^2\big)$.
\end{lemma}
\begin{lemma}\cite{S}
	Let $T$ be a square in $\mathbb{R}^2$ and $h$ be a smooth function such that $Supp(h)\subseteq int(T)$. Let $\Delta_T$ be the Laplacian on $T$. Then, for ${\nu}>0,$
	$Tr_{w}\big(M_h(-\Delta_T + {\nu})\big)^{-1}=\pi\,\int\,h(u)du.$
\end{lemma}
\begin{proposition}
	Let $\Delta^m$ be the magnetic Laplacian and $M_g$ be the multiplication operator by the function $g$. Then,  {for\, $\nu>0,$}
	\begin{enumerate}
		\item $M_g\big(-\Delta^m + {\nu}\big)^{-d}$ is of Dixmier trace class.
		\item $Tr_{w}\big(M_g(-\Delta^m + {\nu})\big)^{-d}= \pi^d {v^m}(g).$
	\end{enumerate}
\end{proposition}
{\it Proof:} Let $h\in Dom(\Delta^m)\subseteq L^2\big(\mathbb{R}^2\big).$ We get $gh\in Dom(\Delta_T)$ and according to Lemma \ref{l1}, we have \begin{eqnarray}
\big(\Delta^m_T\,M_g - M_g\,\Delta^m\big)(h)
&=& \big(\Delta_T\,M_g - M_g\,\Delta \big)(h)
+ \big( T^0_{jk}\,M_g -  M_g\,T^0_{jk}\big)(h).
\end{eqnarray}
Using the Proof of Theorem $5.2$\cite{CGS}, we have
\begin{equation}
\big(\Delta^m_T\,M_g - M_g\,\Delta^m_T\big)(h) = Bh + Ah,
\end{equation}
where $B$ is given as in  \cite{CGS}, and
\begin{eqnarray}
A &=& -M_{T^0_{jk}}g + 2i\Big(\sum_{j=1}^{2}\pi\,iG_jM_{\partial_j}g + 2\pi^2G^2_j \Big)\nonumber\\
&+& 2\,ig^{jk}\sum_{j\neq k}\big(M_{\partial_j}g - 2\pi\,iG_j\big).\big(\partial_k - 2\pi\,iG_k\big).
\end{eqnarray}
From the relation $(5.2)$\cite{CGS}, we have
\begin{eqnarray}
M_g(-\Delta^m + {\nu})^{-1} - (-\Delta^m_T + {\nu})^{-1}M_g 
&=& (-\Delta_T+{\nu})^{-1}B(-\Delta+{\nu})^{-1}\nonumber\\ &+& [ T^0_{jk} , M_g](h). 
\end{eqnarray}
Using the fact that the derivations $d_j$ for $j\in\{1,2\}$ commute, we obtain 
\begin{equation}
M_g(-\Delta^m + {\nu})^{-1} - (-\Delta^m_T + {\nu})^{-1}M_g =  (-\Delta_T+{\nu})^{-1}B(-\Delta+{\nu})^{-1}.
\end{equation}
The result is obtained following step by step \cite{CGS}. $\cqfd$

Let us consider now the noncommutative case. Using the Lemma \ref{l1} in Section $2$, we get the following definition:
\begin{definition}
	The magnetic Lindbladian $\mathcal{L}_0^m$ generated by the magnetic derivation on $\mathcal{B}$ is defined as follows:
	\begin{equation}
	\mathcal{L}_0^m(a(g)) = 	\mathcal{L}_0(a(g)) + a(T^0_{jk}g),
	\end{equation}
	where $(j,k)\in\{1,2\}$,  $g\in\mathcal{C}^{\infty}_c\big(\mathbb{R}^{2d}\big)$ and $	\mathcal{L}_0$ is the Lindbladian on $\mathbb{R}^{2d}$.
\end{definition}

Since $\mathcal{L}^m_0 = \Delta + T^0_{jk}$ and the operator $T^0_{jk}$ is self-adjoint in $L^2\big(\mathbb{R}^{2d}\big),$
according to \cite{CGS}, the operator $\Delta$ has a self-adjoint extension in $L^2\big(\mathbb{R}^{2d}\big)$. We deduce that $\mathcal{L}^m_0$ is a self-adjoint operator in the same space. Thus, in this case, the magnetic heat semigroup is given by $\mathcal{T}^m_t = e^{t\mathcal{L}^m_0}$ , and using the Lemma and the fact that the operators are non commuting, we have $\mathcal{T}^m_t = \mathcal{T}_te^{t\,T^0_{jk}}$ with $(j,k)\in\{1,2\}$. Similarly to the previous sections, we can define the volume form on $\mathcal{B}^{\infty}$ as
${v^m}(a(g)) = \displaystyle\lim_{t\longrightarrow 0^+}t^dTr\big(a(g)\mathcal{T}^m_t \big)$. We therefore get:
\begin{proposition}
	The volume form is given by
	\begin{equation}
	{v^m}\big(a(g)\big) = f(G_1,G_2)\int g\,du
	\end{equation}
\end{proposition}
{\it Proof:} It is similar to the Proof of the Proposition \ref{p2}. The magnetic integral operator $a(g)\mathcal{T}_t^m$ in the Hilbert space $\mathcal{H}$ has the kernel given by
\begin{equation}
\tilde{K}^m_t(u,v)=\hat{g}(u-v)e^{-(t/2)w(v)}e^{ip(u,v)/2}.
\end{equation}
It is obvious to see that $\tilde{K}^m_t(u,u)= K^m_t(u,u)$. As $K^m_t$ is continuous in $\mathbb{R}^{2d}$, we deduce the continuity of $\tilde{K}^m_t$ in the same space.  Since 
\begin{eqnarray}
Tr\big(a(g)\mathcal{T}_t^m\big)
&=&f(G_1,G_2)Tr\big(a(g)\mathcal{T}_t\big),
\end{eqnarray}
using \cite{FG}, we have
\begin{equation}
Tr\big(a(g)\mathcal{T}_t^m\big)= (2\pi)^d\,t^{-d} f(G_1,G_2).
\end{equation}
Thus, we get the result.
$\cqfd$
\subsection{Magnetic quantum stochastic process}
The stochastic process associated with the heat semigroup in the case of classical (or noncommutative $C^{*}$-algebra) of $\mathcal{C}_{0}(\mathbb{R}^{2d})$ is the well known standard Brownian motion \cite{S}. For the noncommutative $C^{*}$-algebra $\mathcal{A}$, Sinha and co-workers  \cite{CGS} studied the stochastic process associated with the heat semigroup in the case of $\mathcal{B}(L^{2}(\mathbb{R}))$ by the Stone-von Neumann theorem on the representation of the Weyl relation \cite{FG}: 

\begin{equation}\label{r}
(U_{ {\alpha}} f)( {x})=f (  {x} +  {{\mu}} )~~\mbox{,}~~
(V_{ {\mu}}f )(  {x}) = e^{i {\alpha}.{x}} f
( {x}),
\end{equation}

with $ q_k,p_k  ( k=1,2 \ldots, n)$ playing the role of  generators of $ V_{ {\mu}} $ and $ U_{ \alpha},$ respectively.  They are the position and momentum operators in the  Schrodinger representation. {We consider the case $d=1$ and the following quantum stochastic differential equation
 in $ L^2 ({\mathbb{R}})
 \otimes \Gamma (L^2({\mathbb{R}}_+,C^2))$\cite{CGS}:
 \begin{equation}\label {ne1a}
 \left \{
 \begin{array}{l}
 dX_t= X_t [ -i p ~dw_1  (t)   -i q~dw_2(t) 
 - \frac{1}{2}\big(p^2 + q^2 \big)
  dt ] ,\\
  X_0=I
  \end{array}
  \right.
 \end{equation}
 where $w_1,w_2$ are independent  Brownian motions.
The next Corollary gives the magnetic Laplacian in the noncommutative $2-$ dimmensional space.
\begin{corollary}\label{co}
	Let $q$ and $p$ be the magnetic momentum and position operators, respectively. Then, the unperturbed noncommutative magnetic Laplacian is  given by:
	\begin{eqnarray}\label{e2}
	\mathcal{L}_0^m &=& \mathcal{L}_0 +1/2\sum_{j=1}^{2}\Big( p^2_j + 2\pi G_jq_j - 4\pi^2G^2_j\Big)\nonumber\\
	&-& \sum_{j\neq k} g^{jk} (q_j-2\pi G_j)(q_k - 2\pi G_k) . 
	\end{eqnarray}
\end{corollary}
{\it Proof:} It is straighforward.
$\cqfd$
 
In this work, let us consider the magnetic quantum stochastic differential equation (m.q.s.d.e) in $ L^2 (\mathbb{R})
\otimes \Gamma (L^2(\mathbb{R}_+,C^2)):$
\begin{equation}\label{ne1}
\left \{
\begin{array}{l}
\displaystyle
dY_t= dX_t + Y_t \tilde{T}_{jk}dt,
\, (j,k)\in\{1,2\} \\
Y_0=I,
\end{array}
\right .
\end{equation}
where $dX_t$ is the quantum stochastic differential equation {(\ref{ne1a})}.

\begin{theorem}
	\begin{enumerate}
		\item The magnetic quantum stochastic differential equation (m.q.s.d.e) (\ref{ne1}) has a unique unitary solution.
		\item If we set $h_t(y)= Y_t(y \otimes I_t)Y_t^* ,$ then for all $ y \in \mathcal{D}_{\theta}^\infty $, $ h_t$ satisfies the magnetic quantum stochastic differential equation (m.q.s.d.e) :
		\begin{equation}
		dh_t(y) = h_t\big(-i[p,y]\big)dw_1(t) + h_t\big(-i[q,y]\big)dw_2(t) + h_t\big( \mathcal{L}^m(y)\big)dt.
		\end{equation}
		Furthermore,
		\begin{equation}\label{me}
		dh_t(y)  = dj_t(y) + h_t(\hat{R}_{jk}(y))dt \quad\mbox{,}\quad (j,k)\in\{1,2\},
		\end{equation}
		where 
		\begin{equation}
		\hat{R}_{jk}(y) = \tilde{T}_{jk}y + y\tilde{T}_{jk}.
		\end{equation}
		Moreover, 
		\begin{equation}
		Eh_t(y) = e^{t\mathcal{L}}(y) + H_t(\hat{R}_{jk}(y)),
		\end{equation}
		where $H_t$ is a primitive of $h_t$.
	\end{enumerate}
\end{theorem}
\textit{Proof:}

\begin{enumerate}
	\item According to \cite{CGS}, the quantum stochastic differential equation $dX_t$ has a unique unitary solution. Then, we deduce that  $dY_t$ has a unique unitary solution.
	\item
	Using It$\hat{o}'s$ formula and table \cite{EH,EHM}, we have
	\begin{equation}\label{ab}
	dh_t(y) = h_t\Big(-i [p,y]\Big)~dw_1(t)+ h_t\Big(-i [q,y]\Big)~dw_2(t) + h_t\Big(\mathcal{L}^m (y)\Big)dt,
	\end{equation}
	where
	\begin{equation}
	\mathcal{L}^m (y)= \mathcal{L}^m_0~y + y~\mathcal{L}^m_0 + y(p^2 + q^2).
	\end{equation}
	Moreover, $\mathcal{L}^m_0 = -\frac{1}{2}(p^2 + q^2) + \tilde{T}_{jk}$ and 
	according to \cite{CGS},   
	the equation (\ref{me}) yields.
	
	The integral form of equation (\ref{me}) implies
	\begin{equation}\label{hj}
	h_t(y) = j_t(y) + H_t(\hat{R}_{jk}(y)).
	\end{equation}
	Performing the same computation as in Section $1$, we get the vacuum expectation.
	
	Using the linearity of vacuum expectation, and \cite{CGS}, 
	we obtain the result.
	$\cqfd$
\end{enumerate}
\subsection{Properties of the mqsde solution}
The goal of this section is to derive differents moments (first, second,  and $r^{th}$ moments) and the variance of the mqsde
solution 
 (\ref{me}). We start  by recalling  the following  Proposition, while the  main result is contained in the theorem below.
\begin{proposition}\cite{PK}
	For every $X\in\mathcal{B}_{0}$, there exists a sequence $\{v^{(n)}_{t}(X)\}$, $t\geq 0$  of  $\left( \xi,v_{0},\mathcal{M}\right)$-adapted processes satisfying \begin{enumerate}
		\item $v^{(0)}_{t}(X)= X,$
		\item $v^{(n)}_{t}(X) = X + \displaystyle\int_{0}^{t}\sum_{i,j\geq 0}v^{(n-1)}_{s}(\theta^{i}_{v}(X))d\Lambda_{i}^{v}(s).$
	\end{enumerate}
\end{proposition}
\begin{theorem}
	Let $h_{t}(x)$ be the solution of the magnetic quantum  stochastic differential equation (\ref{me}) and $j_t(x)$ the solution of the quantum stochastic differential equation {(\ref{e20})}. Then, for $ x\in\mathcal{D}^{\infty}_{\theta},$
		the $r^{th}$-moment and the variance are  given by: 
		\begin{equation}\label{c}
		\left \{
		\begin{array}{l}
		\displaystyle
		Eh^{(r)}_t(x) = Eh_0(x) + \int_{0}^{t}Eh^{(r-1)}_s(\mathcal{L}^m(x) )ds,\\
		\\
		Varh_{t}(x) = (\mathcal{L}^m)^{-1}(x)e^{t\mathcal{L}^m}.\mathcal{L}^m(x)\left(1 - e^{-t\mathcal{L}^m} - e^{t\mathcal{L}^m}(x) \right).
		\end{array}
		\right .
		\end{equation}
	In addition, for $(j,k)\in\{
		1,2\}$, we get 
		
		\begin{equation}\label{d}
		\left \{
		\begin{array}{l}
		\displaystyle
		Eh^{(r)}_t(x) = Ej^{(r)}_t(x) + \int_{0}^{t}Eu^{(r-1)}_s(\tilde{T}_{jk} )ds,\\
		\\
		Varh_{t}(x) = Varj_t(x) - m(G_{t}(\tilde{T}_{jk}(x))) ,~~\mathcal{D}^{\infty}_{\theta},
		\end{array}
		\right .
		\end{equation}
		with
		\begin{equation}\label{e}
		m(y)= y^2 + 2y(Ej_t(x) +1).
		\end{equation}
\end{theorem}	
\textit{Proof:}
	Using the integral form of (\ref{me}) ,  the linearity of the vacuum expectation, 
and the following relation
	\begin{equation}\label{h}
	\int_{0}^{t}Eh_s(-i\sum_{k=1}^{2}[p_k,x])dw_{2k-1}(s) = \int_{0}^{t} Eh_s(-i\sum_{k=1}^{2}[q_k,x])dw_{2k}(s)=0,
	\end{equation}
we have
	\begin{equation}\label{i}
	Eh_t(x) = Eh_0(x) + \int_{0}^{t}Eh_s(\mathcal{L}^m(x) )ds.
	\end{equation}
	The solution of the above equation (\ref{i}) is given by:
	\begin{equation}\label{j}
	Eh_t(x) = \exp(t\mathcal{L}^m_0)(x).
	\end{equation}
	
	According to Lemma (\ref{l1}) and \cite{CGS}, we obtain 
	\begin{equation}\label{l}
	h_t(x) = j_t(x) + G_t(\tilde{T}_{jk}(x)),\quad(j,k)\in\{1, 2\},
	\end{equation}
	where $G_t$ is the primitive of $h_t,$ and 
	the expectation of $h_t$ is 
	\begin{eqnarray}
	Eh_t(x) 
	&=& E j_t(x) + G_t(\tilde{T}_{jk}(x)),\quad(j,k)\in\{1,2\}.
	\end{eqnarray}
	According to the above Proposition and the relation (\ref{h}),  we get
	\begin{equation}\label{o}
	Eh^{(r)}_t(x) = Eh_0(x)  + \int_{0}^{t}Eh^{(r-1)}_s(\mathcal{L}^m(x) )ds.
	\end{equation}
	 For $r=2$ and once (\ref{h}),  we have
	\begin{equation}\label{r}
	Eh^{(2)}_t(x) = Eh_0(x) +  \int_{0}^{t}Eh^{(1)}_s(\mathcal{L}^m(x) )ds,
	\end{equation}
	and the relation (\ref{i}) implies
$	Eh^{(2)}_t(x) =\mathcal{L}^m{^{-1}}(x)e^{t\mathcal{L}^m}(\mathcal{L}^m(x)) - I. $
	By definition of the variance of $h_t$ and after calculation, 
we obtain
	 the second equation of the system (\ref{c}).
	
	Besides, using the relation between $\mathcal{L}_0$ and $T^0_{jk}$, we obtain
	\begin{equation}\label{u}
	h^{(r)}_t(x) = j^{(r)}_t(x) + \int_{0}^{t}h^{(r-1)}_s(\tilde{T}_{jk}(x))ds,
	\end{equation}
	and the linearity of the vacuum expectation gives
	\begin{equation}\label{mg1}
	Eh^{(r)}_t(x) = Ej^{(r)}_t(x) +  \int_{0}^{t}Eh^{(r-1)}_s(\tilde{T}_{jk}(x))ds.
	\end{equation}
	For $r=2$, we obtain  $Eh^{(2)}_t(x) = Ej^{(2)}_t(x)    + G_t(\tilde{T}_{jk}(x)).$
	Thus, \begin{eqnarray}
	Varh_t(x) 
	&=& Varj_t(x) - G_t(\tilde{T}_{jk}(x))^2 - 2Ej_t(x)G_t(\tilde{T}_{jk}(x))\nonumber\\
	&-& G_t(\tilde{T}_{jk}(x)).
	\end{eqnarray}
	Setting $m(y)= y^2 + 2y(Ej_t(x)+ 1)$ yields the result.
$\cqfd$
\section{ Concluding Remarks}
The magnetic  quantum stochastic differential equation associated to the noncommutative magnetic Laplacian in a noncommutative $2-$ torus has been derived. It has been shown that the volume and the volume form of a noncommutative $2-$ torus remain invariant  under a perturbation by inner derivation of the noncommutative magnetic Laplacian.
The vacuum expectation of the solution to the magnetic quantum stochastic differential equation in a noncommutative $2d$-dimensional space has also been computed and discussed.
\section*{Acknowledgments} This work is supported by TWAS Research Grant RGA No. 17-542 RG/MATHS/AF/AC\_G -FR3240300147. The ICMPA-UNESCO Chair is in partnership with Daniel Iagolnitzer Foundation (DIF), France, and the Association pour la Promotion Scientifique de l'Afrique (APSA), supporting the development of mathematical physics in Africa.

\end{document}